\documentclass[aps,pra,reprint,showpacs,longbibliography,superscriptaddress, citeonline]{revtex4-1}
\usepackage{amsmath,amssymb,bm}
\usepackage{graphicx}
\usepackage{latexsym}
\usepackage{color}
\usepackage{hyperref}
\usepackage{placeins}

\newcommand{\linarite}{PbCuSO$_4$(OH)$_2$}

\newcommand{\aver}[1]{\left\langle #1 \right\rangle}

\begin{document}
\title{Multiferroic phases of the frustrated quantum spin-chain compound linarite}

\author{K.~Yu.~Povarov}
    \email{povarovk@phys.ethz.ch}
    \affiliation{Neutron Scattering and Magnetism, Laboratory for Solid State Physics, ETH Z\"{u}rich, Switzerland}
    \homepage{http://www.neutron.ethz.ch/}

\author{Y.~Feng}
   \affiliation{Neutron Scattering and Magnetism, Laboratory for Solid State Physics, ETH Z\"{u}rich, Switzerland}

\author{A.~Zheludev}
    \affiliation{Neutron Scattering and Magnetism, Laboratory for Solid State Physics, ETH Z\"{u}rich, Switzerland}

\begin{abstract}
The dielectric properties of the prototypical frustrated
ferromagnetic spin-chain compound PbCuSO$_4$(OH)$_2$ known as
linarite, are studied across its strongly anisotropic magnetic phase
diagram in single crystal samples. The ferroelectric character of
the principal low-field spin spiral phase is confirmed. The measured
polarization is fully consistent with the previously proposed
magnetic structure. Spontaneous polarization is also detected in two
other field-induced phases but in some cases is incompatible with
previously suggested models for the spin arrangement.
\end{abstract}
\maketitle

\section{Introduction}

The improper ferroelectric nature of magnetically ordered phases in
certain antiferromagnetic materials endows them with a rich
phenomenology and potential technological applications. In recent
years it has given rise to the entirely new research area of
\emph{magnetic
multiferroics}~\cite{Khomskii_JMMM_2006_MFreview,CheongMosotovoy_NatMat_2007_NatureReview,Tokura_RevProgPhys_2014_MFreview}.
Correlations between magnetic ordering and dielectric properties
have been found in very different classes of materials. On one end
of the spectrum are rare earth compounds, with a large spin value
and truly dramatic magnetoelectric
effects~\cite{GotoKimura_PRL_2004_GiantMCeffect,KagawaMochizuki_PRL_2009_GiantMCdomains}.
The other limit is $S=1/2$ organometallic cuprates, whose extreme
quantum fluctuations fully suppress long-range order, which is only
restored by a field-induced quantum phase
transition~\cite{Schrettle_PRB_2013_SulMF,PovarovReichert_PRB_2015_Sulf}.
A more conventional type of multiferroic cuprates is systems
featuring edge-sharing copper-oxygen chains. Among these are
LiCu$_2$O$_2$~\cite{Park_PRL_2007_LiCu2O2Multiferr} and
LiCuVO$_4$~\cite{Mourigal_PRB_2011_LiCuVO4spincurrent}, which have
long served as prototype materials for the study of
multiferroicity~\cite{Seki_PRL_2008_LiCu2O2chiralitySwitching,Furukawa_PRL_2010_1Delectromagnon}.
In these systems the source of electric polarization is a
helimagnetic arrangement of spins that breaks inversion
symmetry~\cite{Katsura_PRL_2005_MFmicro,Mostovoy_PRL_2006_MFmacro}.
Helimagnetism, in turn, results from a  geometric frustration of
magnetic interactions. Specifically,  the Heisenberg exchange
constant $J_{1}$ between the nearest-neighbor Cu$^{2+}$ spins
 is ferromagnetic, and competes with the antiferromagnetic next-nearest-neighbor coupling $J_{2}$.

The most recently studied member of the frustrated
copper-oxide-chain family is the natural mineral linarite
\linarite~\cite{Baran_PhysStatSol_2006_LinariteInitiation}. It is
also perhaps the most interesting one: the estimated ratio of
exchange constants $J_{1}/J_{2}\simeq-2.8$ places linarite very
close to the quantum critical point at $J_{1}/J_{2}\simeq-4$, where
the ferromagnetic interaction takes over and the ground state
changes to a fully polarized one. At the same time, the saturation
field in linarite is below $10$~T, making its entire magnetic phase
diagram easily accessible
experimentally~\cite{Schapers_PRB_2013_LinariteBulk}. To date, up to
five distinct magnetic phases have been identified. This complex
behavior emphasizes the highly frustrated nature of this spin system
and the importance of quantum spin
fluctuation~\cite{Schapers_PRB_2013_LinariteBulk,Willenberg_PRL_2016_LinariteSDWs}.
The primary Phase~I, occurring at zero applied magnetic field, has
been identified as a spin spiral. It would be natural to expect this
phase to generate nonzero electric polarization, simply by analogy
to LiCu$_2$O$_2$ and LiCuVO$_4$. Indeed, previous studies have
detected the presence of bias-induced electric polarization in
linarite \emph{powder} samples below
$T_{N}$~\cite{Yasui_JPSJ_2011_LinariteMF}. The challenge, remains to
perform more detailed studies on a \emph{single-crystal} sample, in
order to determine the direction of polarization, its relation to
the underlying magnetic order, and its evolution in external
magnetic fields. This is the issue addressed in the present study.

\begin{figure}
  \includegraphics[width=0.5\textwidth]{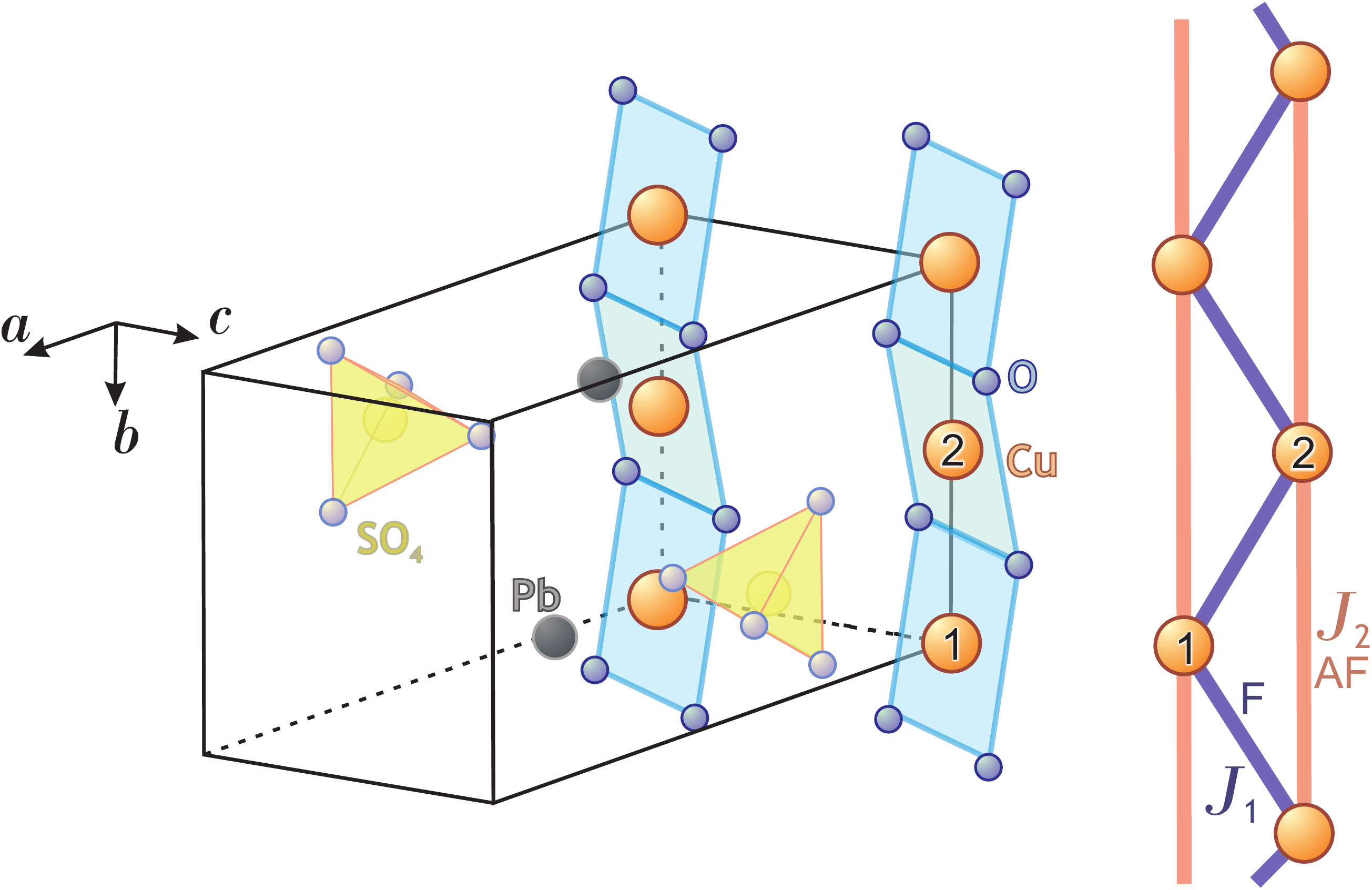}\\
  \caption{Left: Crystal structure of linarite {\linarite}. The two
  copper positions are labeled. Hydrogen atoms adjacent to the in-chain oxygen atoms are omitted for clarity.
   Right: The corresponding diagram of a basic in-chain Heisenberg Hamiltonian with ferromagnetic $J_{1}$ and antiferromagnetic $J_{2}$ interactions (values are given in the text).}\label{FIG:linaritestructure}
\end{figure}

\begin{figure}
  \includegraphics[width=0.5\textwidth]{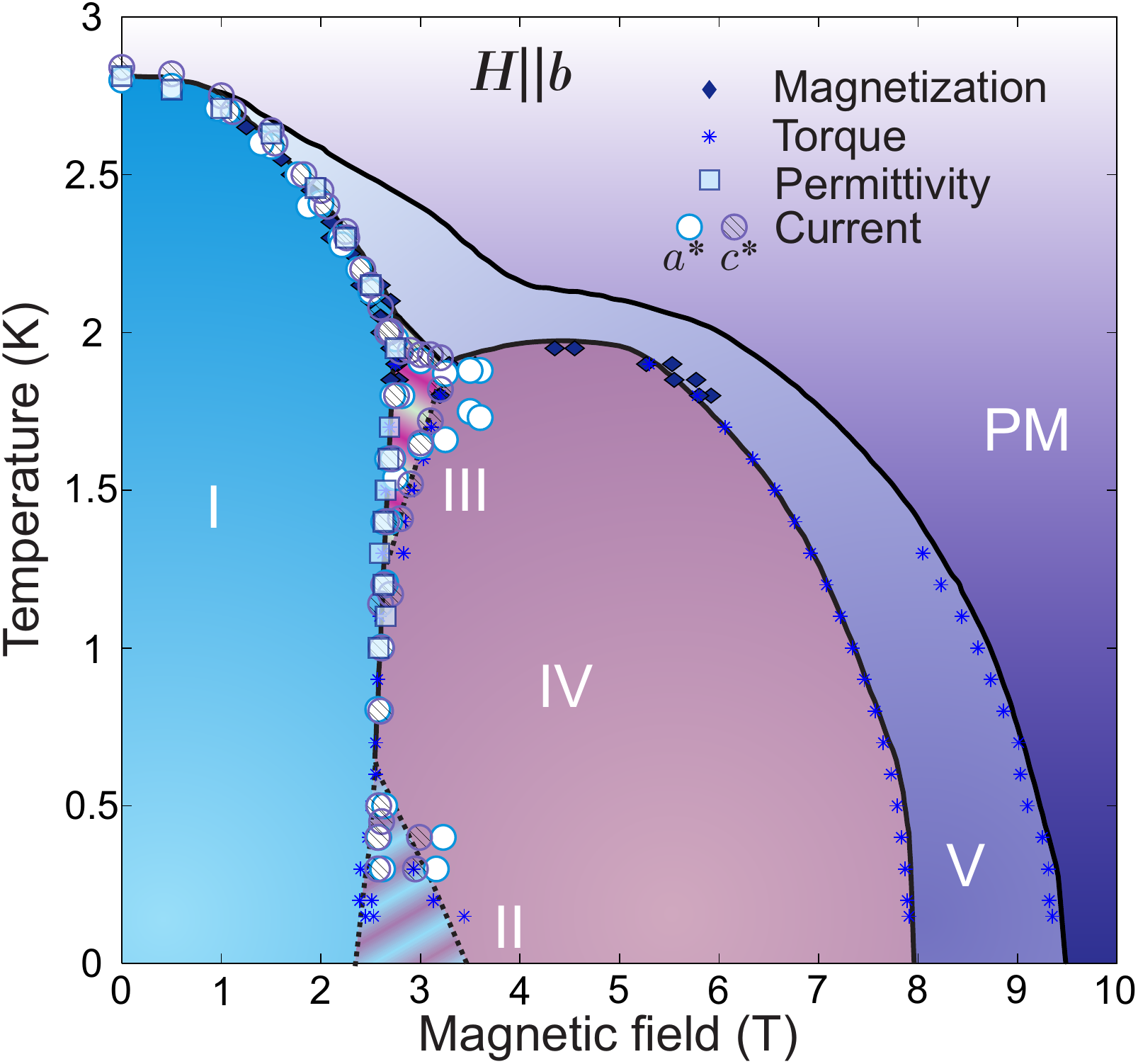}\\
  \caption{Phase diagram of linarite for a field applied along the $\mathbf{b}$ direction.
  Following Ref.~\cite{Willenberg_PRL_2016_LinariteSDWs} the distinct thermodynamic phases are an elliptical spiral (I),
  a collinear N\'{e}el antiferromagnetic phase (IV), a spin-density wave (V), and a circular helix coexisting with collinear ordering (III).
  Phase II is supposed to be a metastable mixture of Phases I and IV. Solid lines are the phase boundaries according to
  Ref.~\cite{Willenberg_PRL_2016_LinariteSDWs}; symbols are the results of this work.}\label{FIG:linaritePhD}
\end{figure}

Blue transparent crystals of linarite belong to the monoclinic
$P2_{1}/m$ (No. 11) space group. The unit cell dimensions are
$a=9.68$, $b=5.65$, $c=4.68$~\AA, with the angle between $a$ and $c$
being
$\beta=102.6^{\circ}$~\cite{Willenberg_PRL_2012_LinariteFrustrated}.
The structure of the material is shown in
Fig.~\ref{FIG:linaritestructure}. There are two copper atoms per
unit cell. Together with surrounding oxygen ions they form a ribbon
chain of Cu-O plaquettes, propagating along the high-symmetry
$\mathbf{b}$ direction. Magnetization measurements supported by
 first-principle calculations lead to estimates of the main exchange parameters of $J_{1}\simeq-8.6$ and
$J_{2}\simeq 3.1$~meV~\cite{WolterLipps_PRB_2012_LinariteESR}. A
rich magnetic phase diagram (Fig.~\ref{FIG:linaritePhD}) was
revealed below
$T_{N}\simeq2.8$~K~\cite{Schapers_PRB_2013_LinariteBulk,Willenberg_PRL_2016_LinariteSDWs}.
Five distinct magnetic phases exist for
$\mathbf{H}\parallel\mathbf{b}$. The zero-field elliptic spin spiral
(I) changes to a canted commensurate structure (IV) around $3$~T,
with the transition line splitting into regions of phase coexistence
at low and high temperatures. The high-temperature region,
labeled~III, is supposed to be a mixture of Phase IV and alternative
spiral configuration, different from Phase~I. The low-temperature II
region is a metastable mixture of Phases I and IV. Finally, for
$\mathbf{H}\parallel\mathbf{b}$, there is an unusual high-field
phase believed to be a spin-density wave state (V). Remarkably, for
a magnetic field applied transverse to the $\mathbf{b}$ direction,
there appears to be only a single magnetic phase, namely, the spiral
state~I.

\begin{figure}
\begin{center}
  \includegraphics[width=0.4\textwidth]{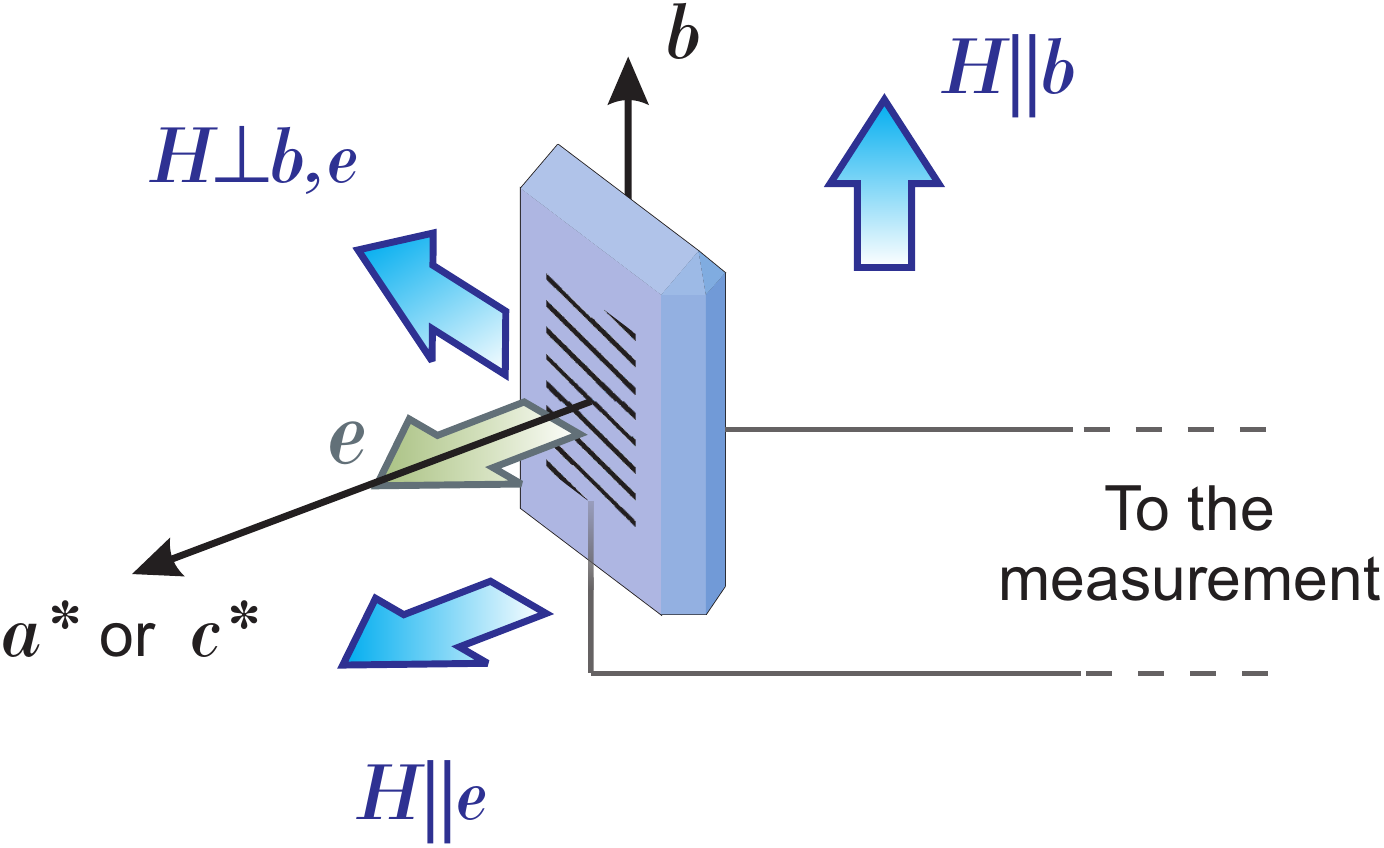}\\
  \caption{Measurement schematics. A small crystal of linarite has electrodes connected to its principal
  surface, either [100] or [001], while the $\mathbf{b}$ axis is in the elongated direction.
  This corresponds to measuring the dielectric properties (in the direction given by $\mathbf{e}$) along either $\mathbf{a}^{\ast}$ or $\mathbf{c}^{\ast}$. Three naturally possible orientations
  of the external magnetic field are also shown.}\label{FIG:geometry}
\end{center}
\end{figure}

\section{Experimental details}

The present study employed natural single crystals of linarite from
the Grand Reef Mine, Arizona, USA. For all samples, the crystal
structure was verified at room temperature by means of x-ray
diffraction (BRUKER APEX II single-crystal diffractometer), and
found to be in good agreement with the previously published
data~\cite{Bachmann_ActCryst_1961_LinariteStructure,Schapers_PRB_2013_LinariteBulk}.
We have also carefully checked the magnetic phase diagram of our
samples and found it to be consistent with previous studies. As will
be reported in detail elsewhere, to that end we used standard SQUID
magnetometry (Quantum Design MPMS) and a home-built cantilever
torque magnetometry setup. The corresponding experimental points,
obtained for several samples, are shown in
Fig.~\ref{FIG:linaritePhD} (rhombi and crosses), in direct
comparison with phase boundaries reported in
Ref.~\cite{Willenberg_PRL_2016_LinariteSDWs} (lines).

For our dielectric experiments we have carefully selected a number
of single crystals in which either the [100] or the [001] faces were
well developed, and where the $\mathbf{b}$ direction could be
clearly identified. The initial choice was based on the
morphology~\footnote{Linarite crystals typically have a prismatic
morphology with elongation along the $b$ axis. In addition to this
easily identifiable direction, typical prism also has the largest
facet corresponding to [100] plane (the main cleavage plane of
linarite). However, there is a reasonable chance of finding a
crystal which has the [001] plane (the second cleavage plane) as the
largest facet instead. These two cases were distinguished at the
next step, where the x-ray diffraction check was performed.}, which
was further verified by x-ray diffraction. The typical area of such
faces was $1-2$~mm$^{2}$, with a typical transverse sample thickness
of about $0.5$~mm. The crystalline plates thus selected were
sandwiched between external field electrodes (see
Fig.~\ref{FIG:geometry}). Correspondingly, the dielectric properties
were probed along either the $\mathbf{a^{\ast}}$~\footnote{As in the
standard reciprocal lattice notation.} or $\mathbf{c^{\ast}}$
directions. For each of these two cases, we used three principal
measurement geometries: $\mathbf{H}\parallel \mathbf{b}$,
$\mathbf{H}\parallel \mathbf{e}$ and $\mathbf{H}\parallel
\mathbf{b}\times\mathbf{e}$, where $\mathbf{e}$ is the direction
transverse to the electrodes. The dielectric permittivity was
measured by an Andeen--Hagerling 2550A capacitance bridge using a
3-terminal scheme. Pyroelectric current measurements were performed
with a Keithley 617A electrometer.

All experiments were carried out in the standard Dilution
Refrigerator inset in the Quantum Design 9~T PPMS. This imposed some
limitations on the types of pyro- or magnetoelectric current scans
that could be performed: measuring current vs. $T$ was feasible only
in the range $1$--$4$~K in the evaporative mode of Dilution
Refrigerator operation. This configuration allowed us to achieve a
good signal-to-noise ratio at a temperature sweep rate of
$0.5$~K/min, while the difference between heating and cooling data
sets remained negligible. In the dilution cooling regime below $1$~K
the increase in thermal coupling times prevented a collection of
meaningful current data in temperature sweeps. In contrast,
measurements of current vs. $H$ were possible at all temperatures
down to approximately $0.2$~K at an optimal sweeping rate of
$0.01$~T/s.

\section{Results}
\subsection{Magnetic field along the $\mathbf{b}$ axis}

\begin{figure}
  \includegraphics[width=0.5\textwidth]{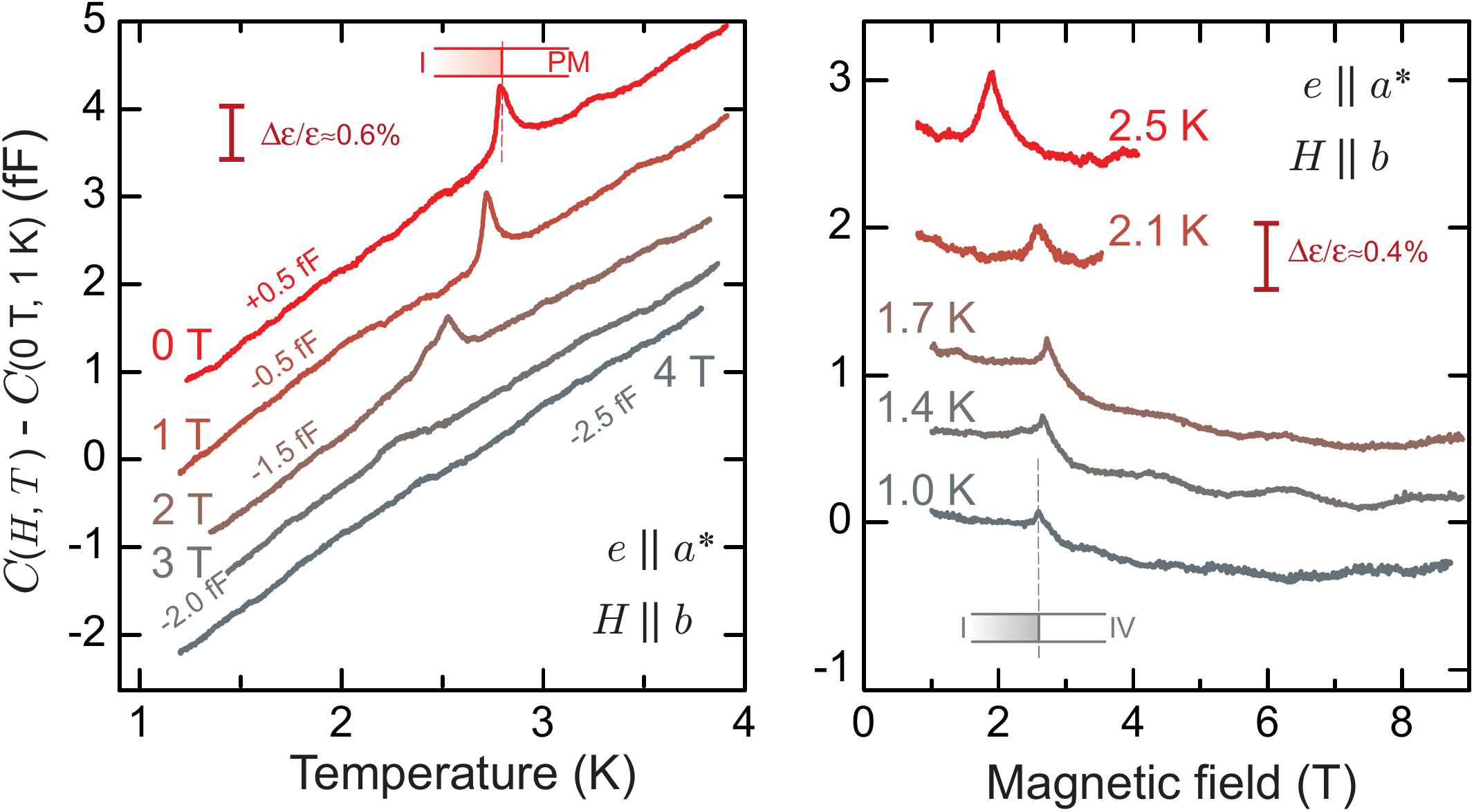}\\
  \caption{Examples of measured $\mathbf{a^{\ast}}$-dielectric permittivity anomalies in
  linarite in a magnetic field along the $\mathbf{b}$ direction. Data are plotted as the difference between the measured capacitance and the reference value observed at $T=1$~K,
  $H=0$~T. In the left panel some constant offsets are additionally introduced for
  clarity.
  Phase boundaries according to Ref.~\cite{Willenberg_PRL_2016_LinariteSDWs} are shown for some of the curves.}\label{FIG:Permit}
\end{figure}

Figure~\ref{FIG:Permit} shows a number of representative dielectric
permittivity scans, measured with the electric field along the
$\mathbf{a^{\ast}}$ direction. In contrast to the previous study by
Yasui~\emph{et al.}~\cite{Yasui_JPSJ_2011_LinariteMF}, we do indeed
find well-defined peaks in the dielectric permittivity occurring at
the boundaries of Phase~I. They are, unfortunately, too weak for a
more quantitative investigation, but serve as markers of the phase
transition and reveal the electrically active nature of Phase~I. In
the phase diagram in Fig.~\ref{FIG:linaritePhD}, their positions are
plotted as squares. No further permittivity anomalies were detected
in magnetic fields exceeding $3$~T. Unfortunately, due to the small
size of the sample the measurement background is non-negligible and
this prevents us from precisely calibrating the vertical scale in
Fig.~\ref{FIG:Permit} in the sample's dielectric permittivity units.
However, from the data we can estimate $\varepsilon\simeq20$ in the
vicinity of $T_N$. This is in agreement with the earlier
measurements by Yasui~\emph{et
al.}~\cite{Yasui_JPSJ_2011_LinariteMF}. The magnitude of the
zero-field anomaly is then estimated as only
$\Delta\varepsilon/\varepsilon\simeq6\cdot10^{-3}$, and it weakens
progressively in the applied magnetic field.

The phenomenology of current anomalies turned out to be much richer
and easier to investigate. An important point is that we found the
charge flow associated with the magnetic ordering to occur
\emph{spontaneously}, without any external voltage applied to the
sample. Furthermore, moderate bias voltages (from +250 to -250~V)
applied to the sample during the cooldown process were able to
change the amount of the accumulated charge by only $\pm 30$\%
without causing a polarization sign reversal. This clearly indicates
the existence of a preferred polarization direction, which is
somewhat surprising for a centrosymmetric space group (such as
$P2_{1}/m$ of linarite). Possible explanations of this finding are
discussed in Sec.~\ref{SEC:ESdiscuss}.

\begin{figure}
  \includegraphics[width=0.5\textwidth]{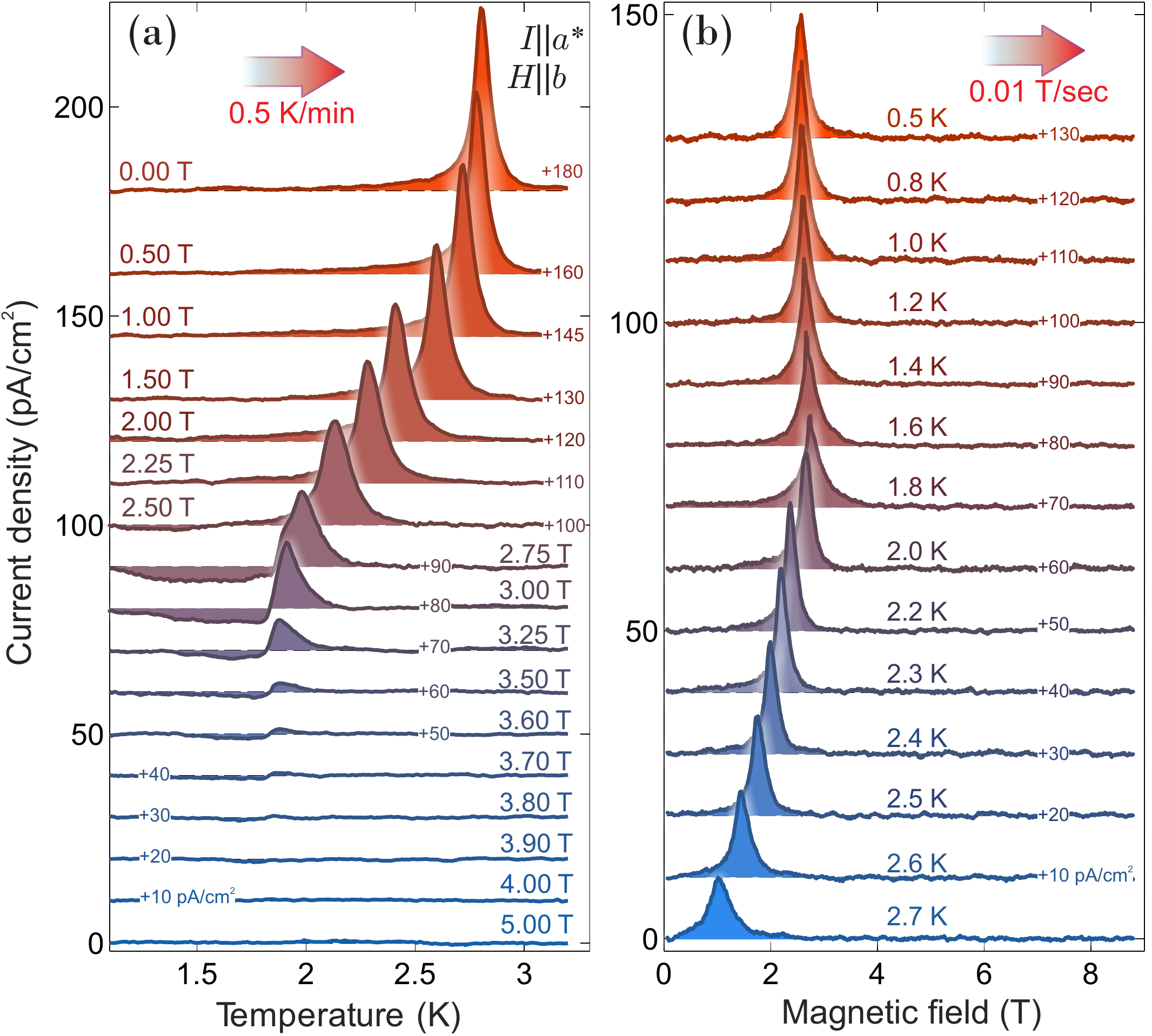}\\
  \caption{Spontaneous current $I_{a^{\ast}}$ through the sample (no bias voltage applied) in a magnetic field applied along the $\mathbf{b}$ direction.
  (a) The pyroelectric current
  as a function of the temperature during the warm-up;
  (b) the magnetoelectric current in an increasing magnetic field. Offsets are explicitly indicated.
  The zero level for each curve is shown by a dashed line.}\label{FIG:IaHb}
\end{figure}

\begin{figure}
  \includegraphics[width=0.4\textwidth]{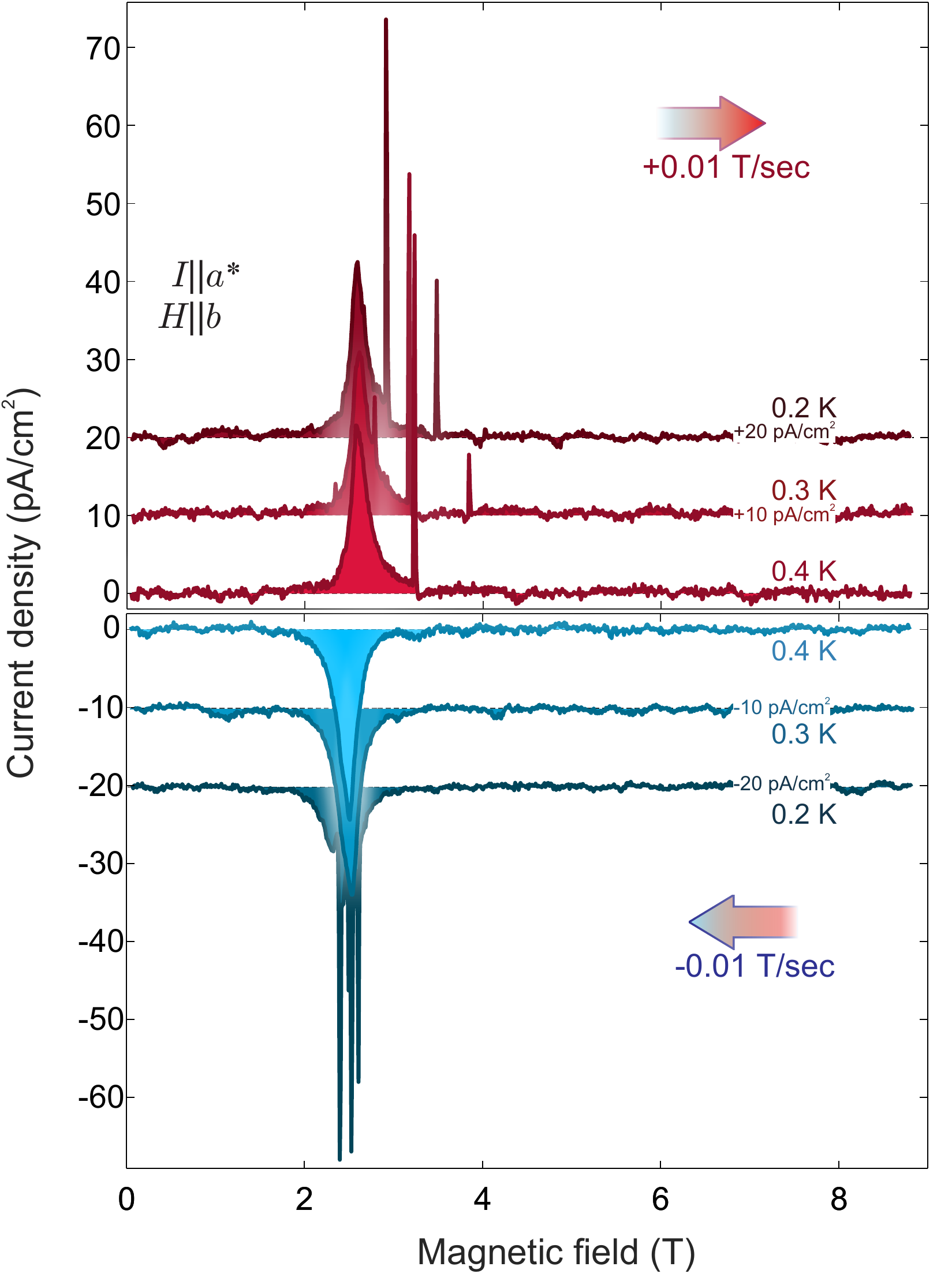}\\
  \caption{Examples of up and down field scans of the $I_{a^{\ast}}$ current below $T=0.5$~K.
  The magnetic field is applied along the $\mathbf{b}$ direction. The difference in the
  scans with increasing vs decreasing $H$, as well as the complex multi-peak response, is related to the metastability of Phase II.}\label{FIG:IaHb_HYST}
\end{figure}

\begin{figure}
  \includegraphics[width=0.5\textwidth]{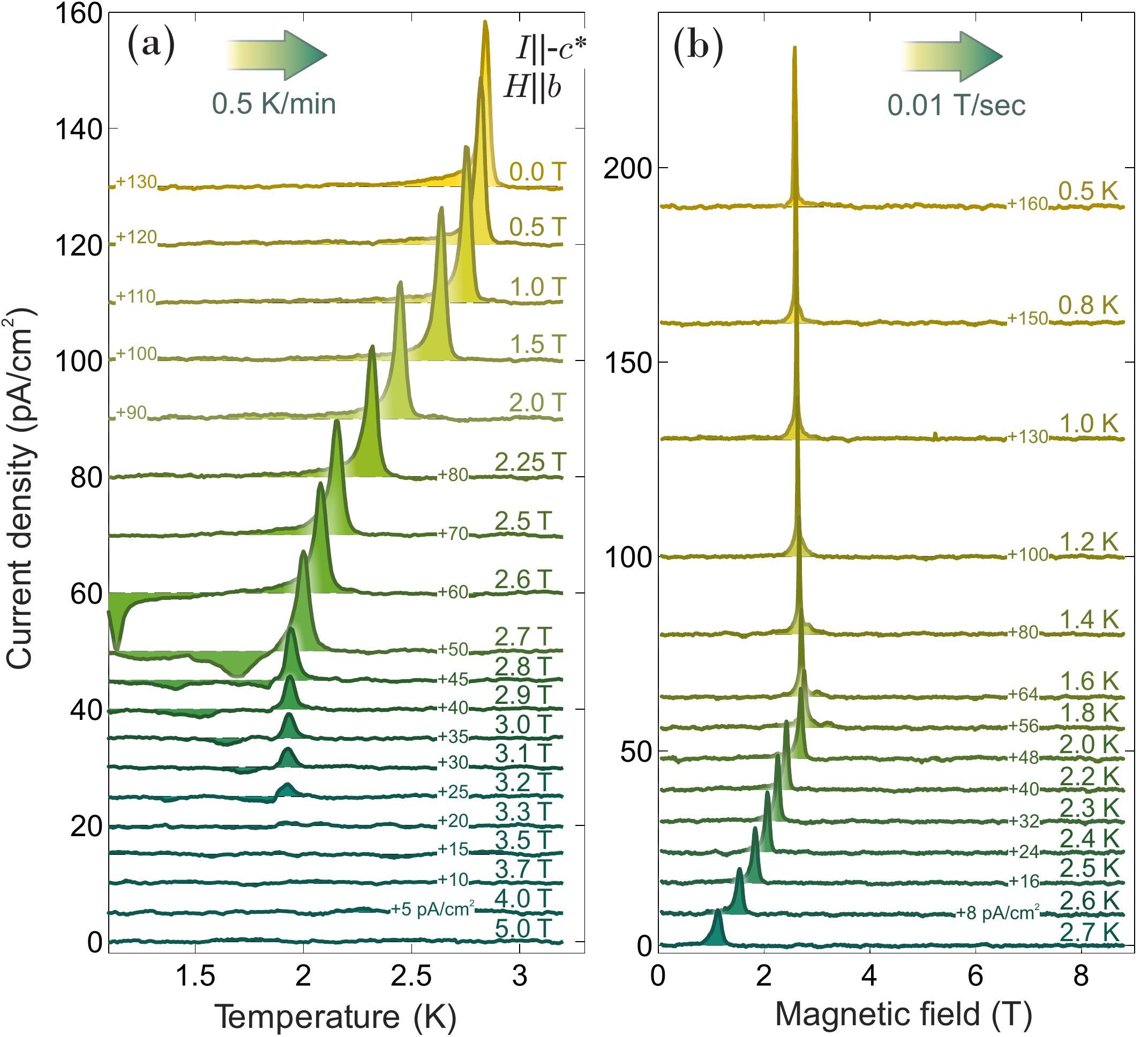}\\
  \caption{The same as Fig.~\ref{FIG:IaHb}, but for the $I_{c^{\ast}}$ component of the current.}\label{FIG:IcHb}
\end{figure}

The pyroelectric current $I_{a^{\ast}}$ measured in temperature
sweeps at several values of applied field is plotted in
Fig.~\ref{FIG:IaHb}(a). In zero magnetic field there is a prominent
peak with a maximum at $T_{N}\simeq2.8$~K. This peak gradually
broadens and shifts to lower temperatures with increasing field up
to approximately $2.5$~T. At this point, upon cooling, one enters
the region labeled III in Fig.~\ref{FIG:linaritePhD}. Here the
pyroelectric current develops a more complex two-peak structure.
First, charge goes into the sample during the cooldown into Phase
III, and then it exits upon further cooling towards the Phase IV.
Above $3.6$~T no transition-related features in the pyroelectric
current could be resolved at these temperatures. This shows the
apparent nonelectric character of Phases IV and V.

Isothermal magnetoelectric current measurements provide a
complementary way of accessing the polarization. The field
dependence of $I_{a^{\ast}}$ is plotted  in Fig.~\ref{FIG:IaHb}(b)
for several temperatures. At first glance there is just a single
peak corresponding to the charge released upon exiting Phase~I.
However, additionally there is a sharp current spike on top of the
broader peak between $0.5$ and $1.5$~K. This marks a polarization
discontinuity along what is a first-order phase transition line. The
discontinuity becomes somewhat softened at the lowest temperatures,
when another region of phase coexistence is approached (Phase II in
the Fig.~\ref{FIG:linaritePhD} phase diagram). As shown in
Fig.~\ref{FIG:IaHb_HYST}, a rather complex behavior emerges below
0.5~K. Upon lowering the temperature the main peak becomes
accompanied by multiple history-dependent sharp satellites. This
directly reflects the metastable nature of Phase
II~\cite{Schapers_PRB_2013_LinariteBulk}. No additional features
could be found at higher magnetic fields at any temperature.

For the $\mathbf{c^{\ast}}$ component of the current the situation
is qualitatively similar. Examples of the $I_{c^{\ast}}$ scans are
shown in Fig.~\ref{FIG:IcHb}. The main difference is the weaker and
much more abrupt character of the corresponding anomalies. The peaks
in $I_{c^{\ast}}$ are much narrower than in $I_{a^{\ast}}$ at
similar temperatures. This is especially pronounced below 1~K at the
first-order transition from Phase~I to Phase IV. At temperatures
below 0.5~K a hysteretic multipeak structure develops, similarly to
that in $I_{a^{\ast}}$ described above.

\subsection{Magnetic field  along the $\mathbf{a^{\ast}}$ and $\mathbf{a}$ directions}

\begin{figure}
  \includegraphics[width=0.5\textwidth]{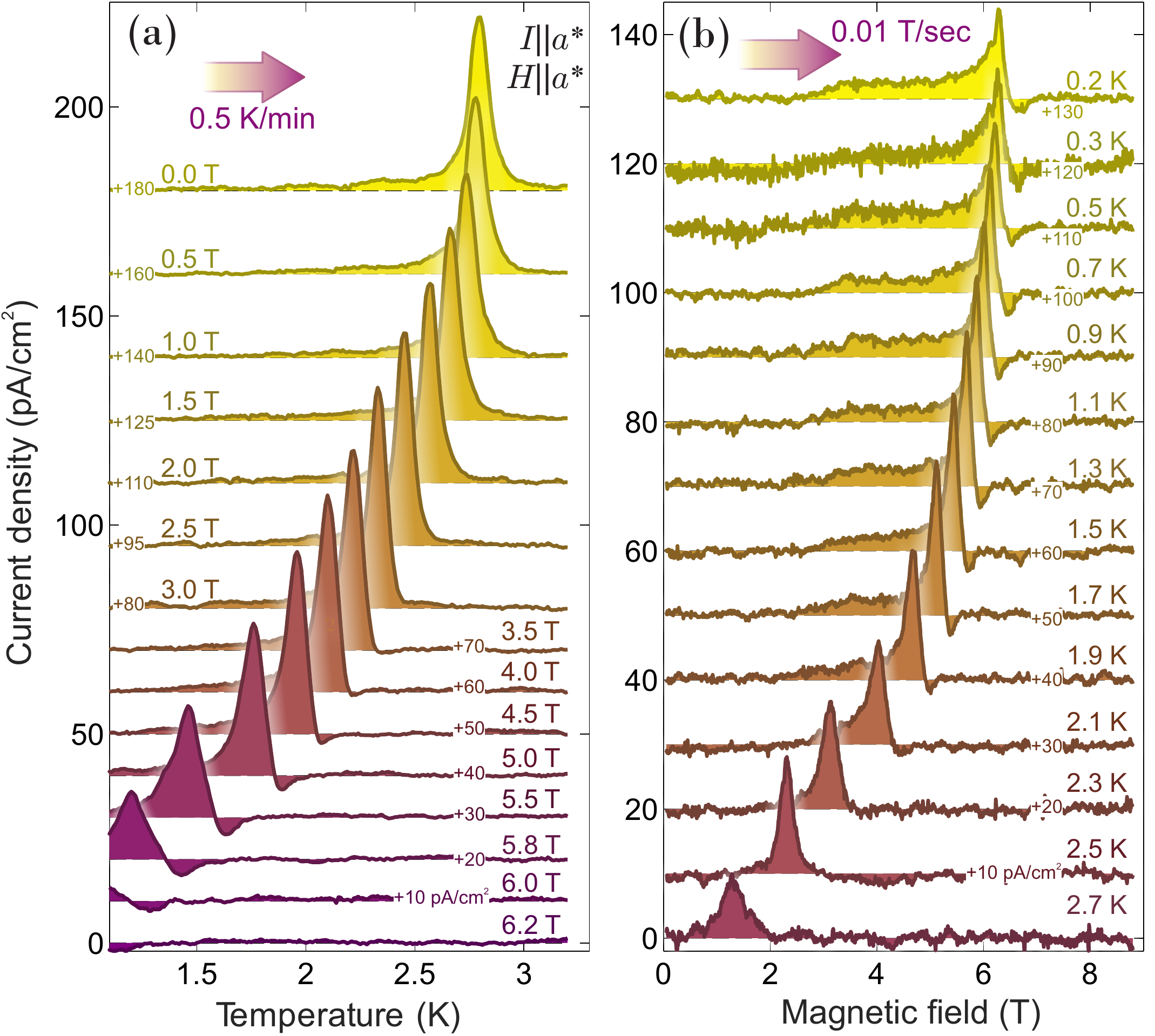}\\
  \caption{The same as Fig.~\ref{FIG:IaHb}, but for the magnetic field $\mathbf{H}\parallel\mathbf{a^{\ast}}$.}\label{FIG:IaHa}
\end{figure}

\begin{figure}
  \includegraphics[width=0.5\textwidth]{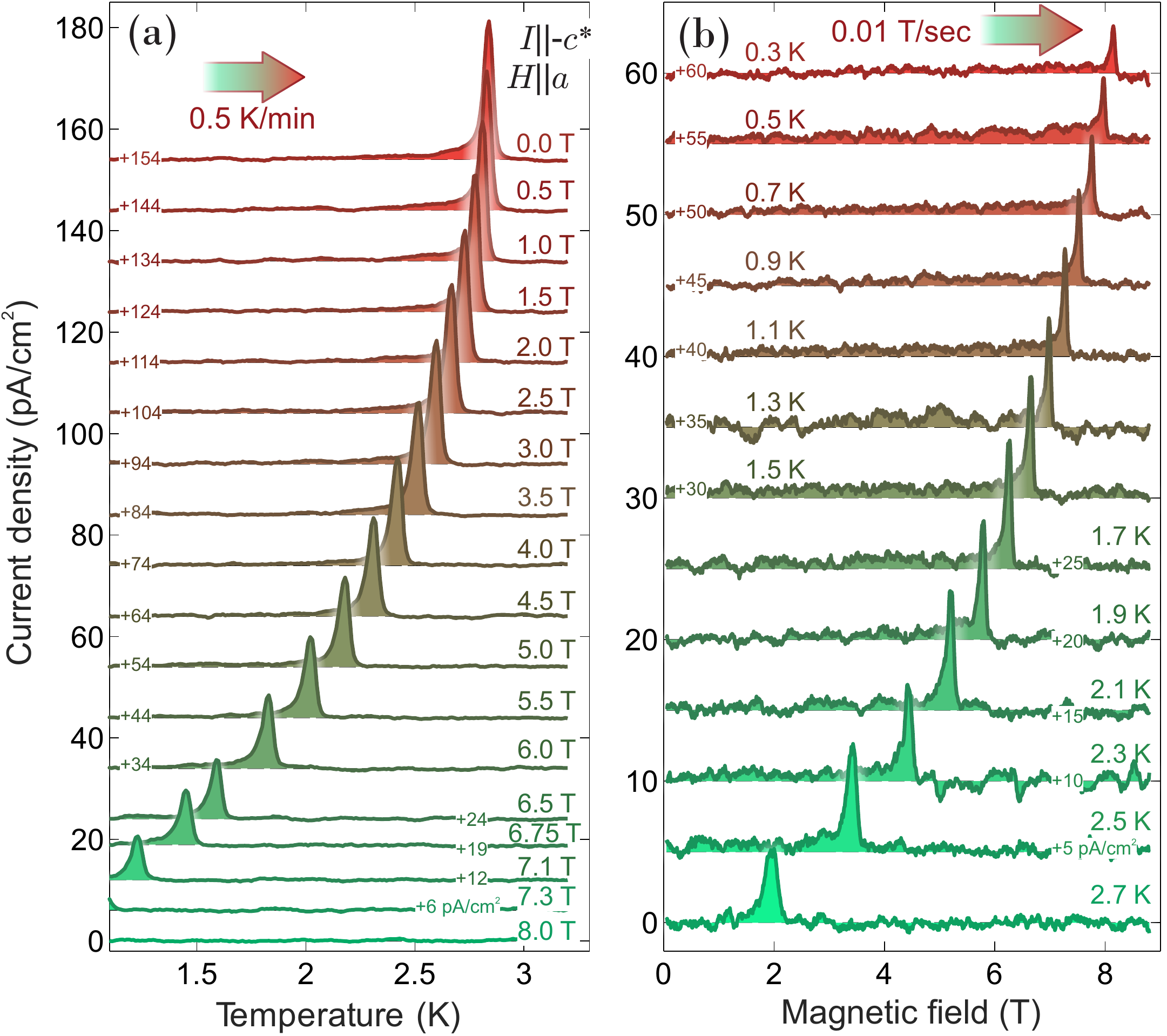}\\
  \caption{The same as Fig.~\ref{FIG:IcHb}, but for the magnetic field $\mathbf{H}\parallel\mathbf{a}$.}\label{FIG:IcHa}
\end{figure}

Anomalies occurring in the current $I_{a^{\ast}}$ as a function of
temperature and magnetic field along the $\mathbf{a}^{\ast}$
direction are shown in Fig.~\ref{FIG:IaHa}. In small fields, the
pyroelectric current behaves in a way  similar to the
$\mathbf{H}\parallel\mathbf{b}$ case. However, at higher fields a
surprising two-peak structure is observed. The reversal of current
direction corresponds to polarization reversal upon cooling. This
behavior is also well pronounced in field scans
[Fig.~\ref{FIG:IaHa}(b)] at lower temperatures. The charge release
is slower than in the $\mathbf{H}\parallel\mathbf{b}$ case. This is
why the amplitude of the $I_{a^{\ast}}$ anomaly is seemingly
reduced. Instead, the transition-related peak has a long tail
stretching to low fields, reflecting the gradual evolution of the
spiral structure towards the full saturation.

The current $I_{c^{\ast}}$ in a magnetic field applied along $\mathbf{a}$
shows slightly different behavior (Fig.~\ref{FIG:IcHa}). We find
only a single sharp peak at all fields and
temperatures. In the low-$T$ regime, it develops an extended tail, similarly to the $I_{a^{\ast}}$ case discussed
above.

\subsection{Magnetic field along the $\mathbf{c}$ and $\mathbf{c^{\ast}}$ directions}

\begin{figure}
  \includegraphics[width=0.5\textwidth]{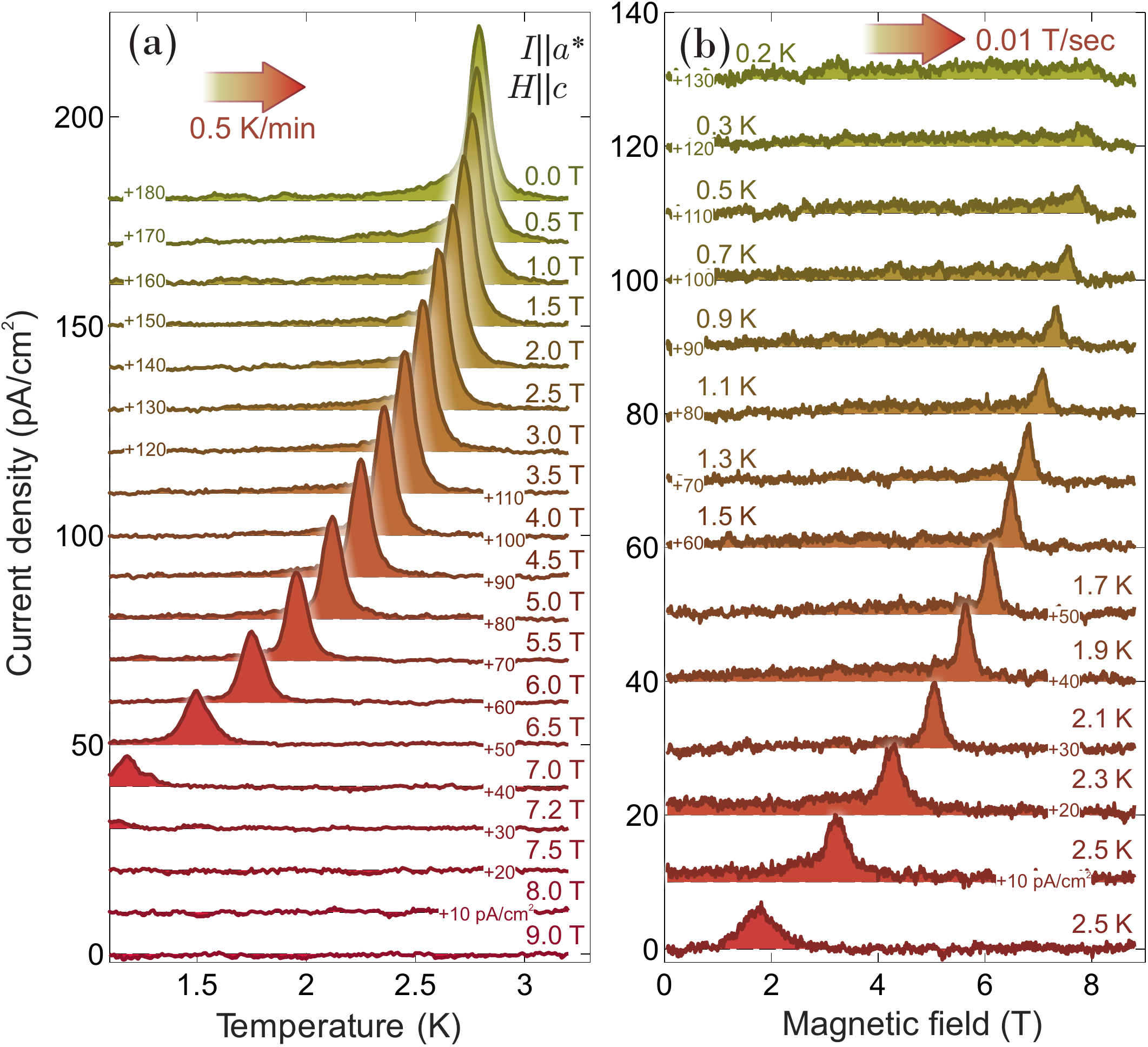}\\
  \caption{The same as Fig.~\ref{FIG:IaHb}, but for the magnetic field $\mathbf{H}\parallel\mathbf{c}$.}\label{FIG:IaHc}
\end{figure}

\begin{figure}
  \includegraphics[width=0.5\textwidth]{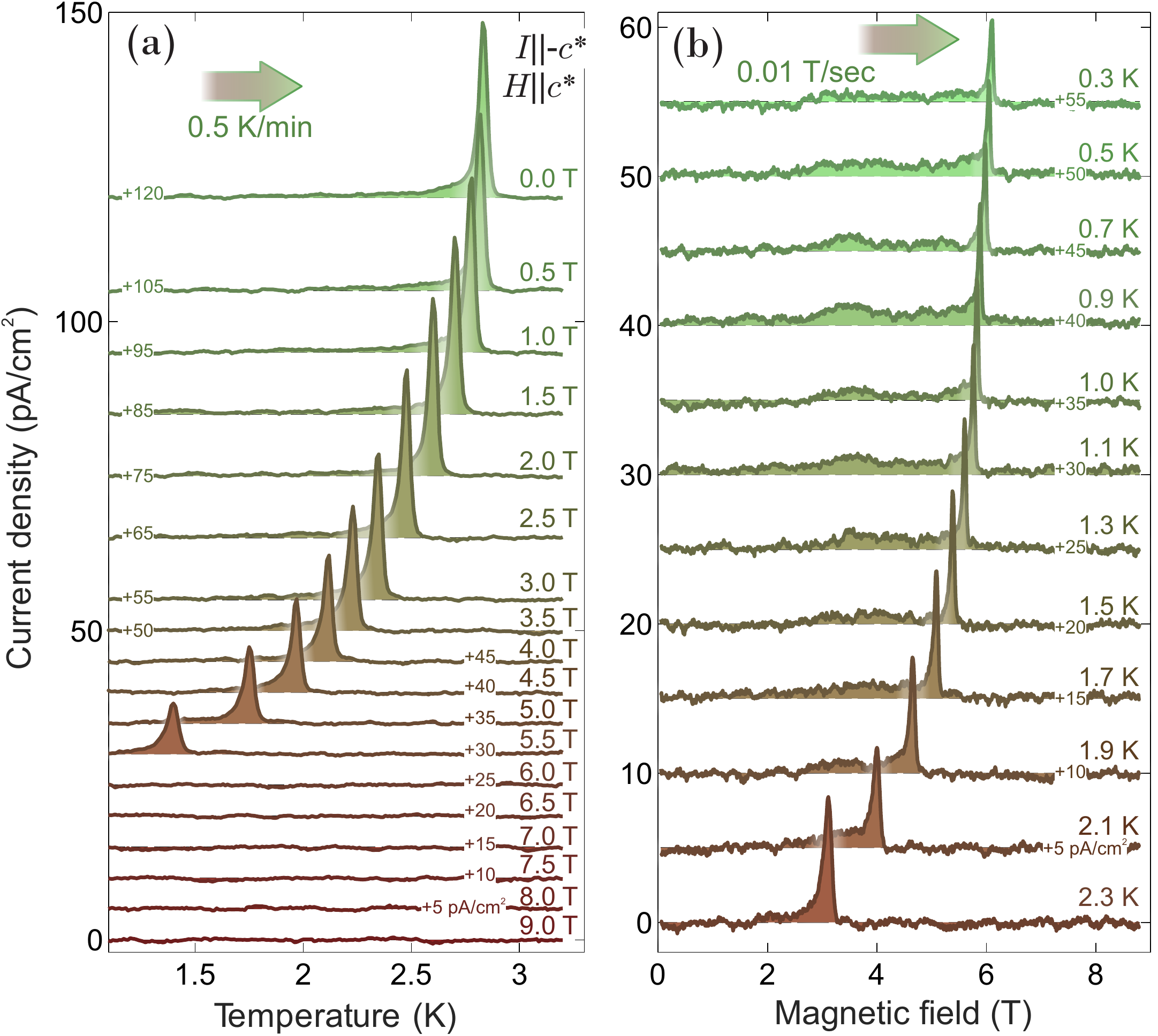}\\
  \caption{The same as Fig.~\ref{FIG:IcHb}, but for the magnetic field $\mathbf{H}\parallel\mathbf{c^{\ast}}$.}\label{FIG:IcHc}
\end{figure}

The cases of a magnetic field applied along the $\mathbf{c}$ and
$\mathbf{c}^\ast$ directions appear to be the most trivial ones.
There is always a single feature in the spontaneous current
occurring at the phase boundary. A summary of $I_{a^{\ast}}$ scans
is present in Fig.~\ref{FIG:IaHc}. Upon lowering the temperature the
peak becomes progressively less pronounced. Below $0.5$~K, instead
of a peak, a small and almost-constant current is detected within
the ordered phase. This corresponds to an almost-constant rate of
charge release.

The $c^{\ast}$ component of electrical current measured in a
magnetic field applied along the $\mathbf{c}^{\ast}$ direction is
plotted in Fig.~\ref{FIG:IcHc}. The observed behavior is almost
identical to that for $\mathbf{H}\parallel\mathbf{a}^{\ast}$.

\section{Discussion}

\subsection{Polarization: A brief summary}

\begin{figure}
  \includegraphics[width=0.5\textwidth]{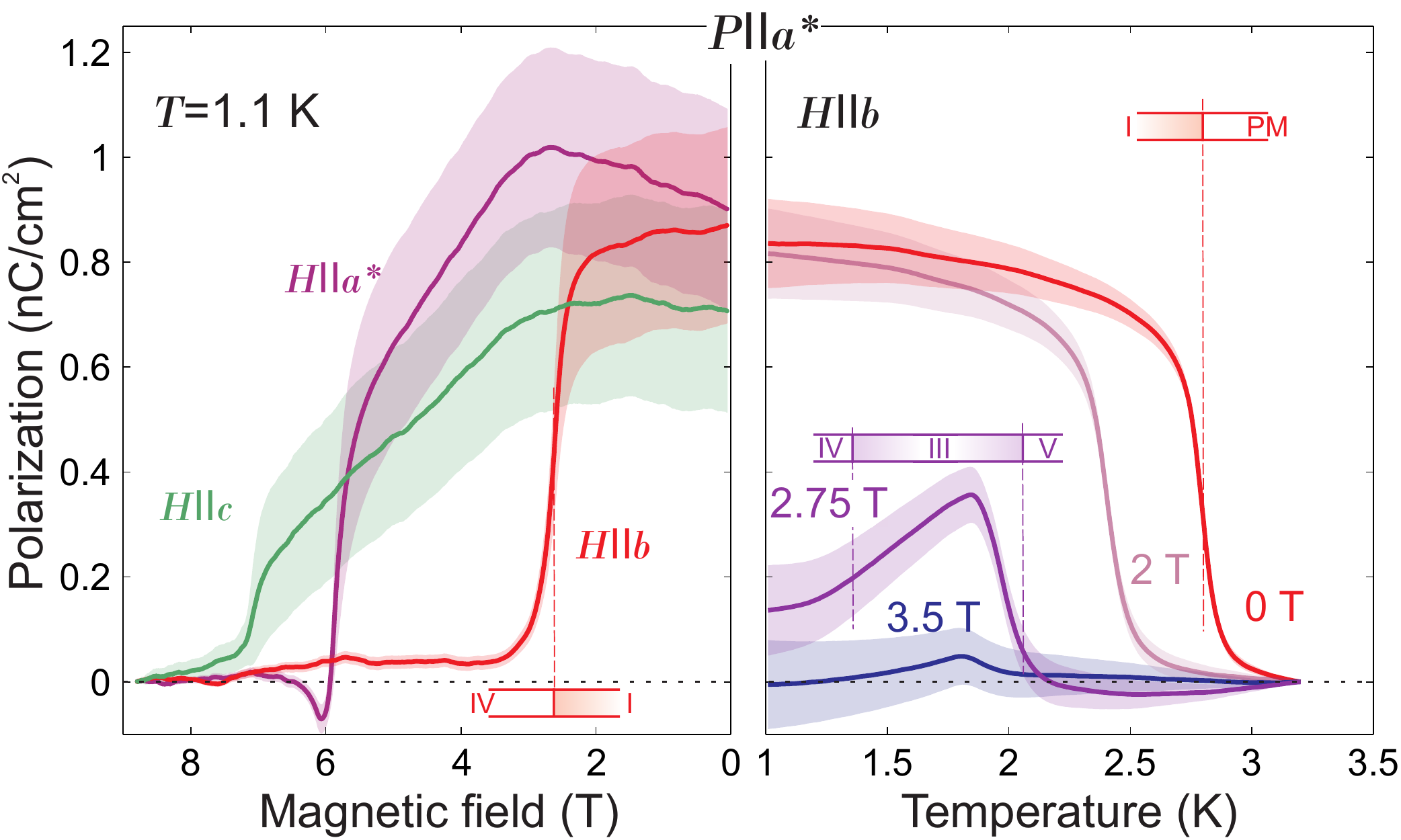}\\
  \includegraphics[width=0.5\textwidth]{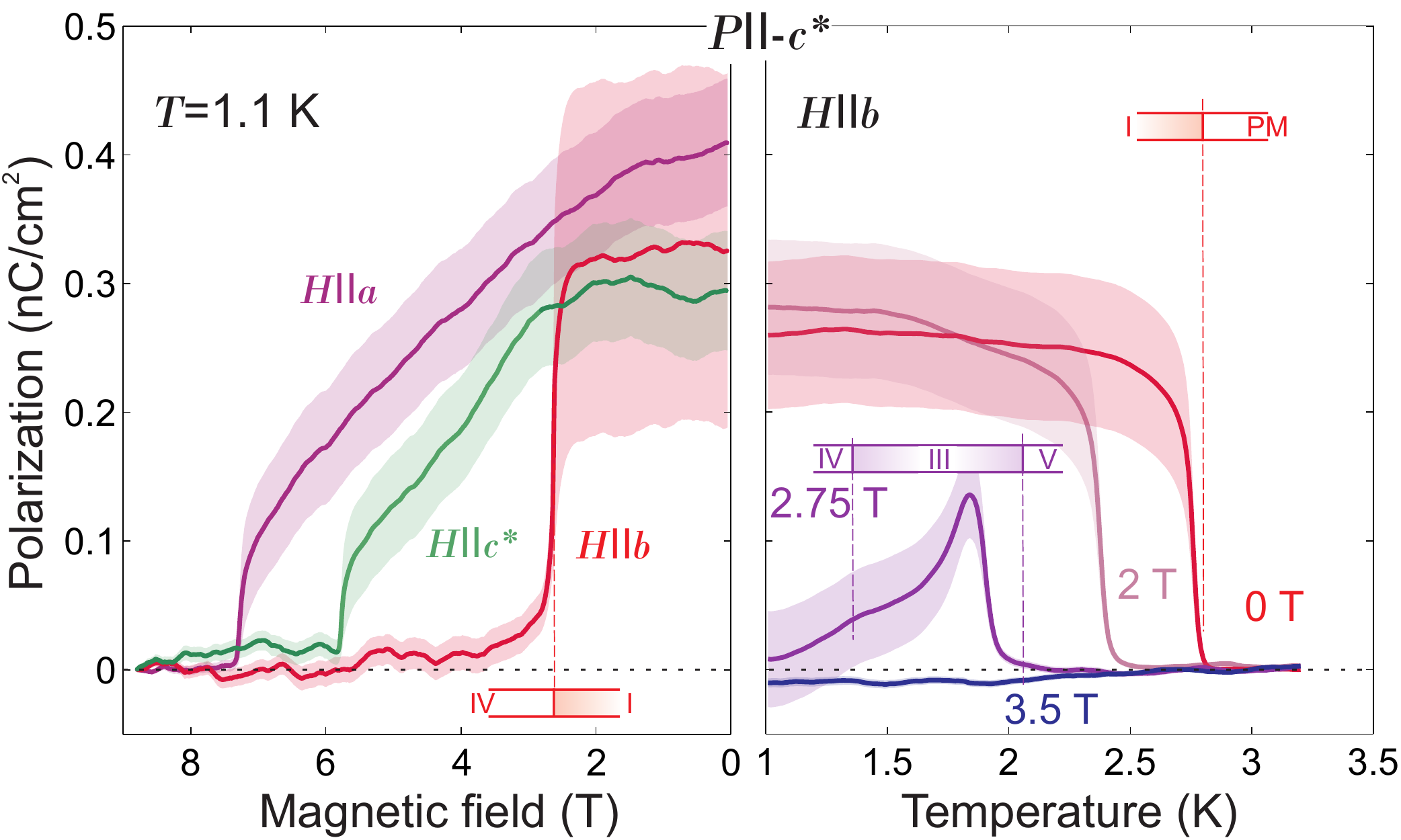}\\
  \caption{Measured evolution of the polarization vector components as functions of the magnetic field and
  temperature. Shaded areas show respective errors accumulating
  during the current integration procedure.
  Phase boundaries according
  to~\cite{Willenberg_PRL_2016_LinariteSDWs} are indicated for
  selected polarization curves.
  }\label{FIG:Pasummary}
\end{figure}

In Fig.~\ref{FIG:Pasummary} we compare representative field and
temperature dependencies of different  electric polarization
components. These plots are deduced from  $T$ and $H$ dependencies
of $I_{a^{\ast}}$ and $I_{c^{\ast}}$, respectively, keeping in mind
that the spontaneous polarization is totally absent at high fields
and high temperatures. In this case the polarization change obtained
by integrating the field or temperature current dependence (always
along the trajectory starting in the paramagnetic phase phase where
$\mathbf{P}=0$) is equivalent to the total polarization.

 Along with the curves one can
also see the associated uncertainty, estimated from the noise level
during the measurement (those shaded areas also happen to be a fair
estimate of the measurement reproducibility). The following
 phenomenology is evident from these data: in small magnetic fields the polarization
emerges at the ordering temperature of Phase  I, and then quickly
saturates. One may argue that both components of $\mathbf{P}(T)$ are
already saturated around 1~K. For a field applied along the
$\mathbf{b}$ axis, the situation drastically changes around $2.5$~T,
where upon cooling one consecutively enters first Phase III, and
then Phase IV. Again, both components of $\mathbf{P}(T)$ rise around
the corresponding transition temperature, but this is followed by a
decrease upon further cooling. The magnitude of the polarization
components is also noticeably reduced compared to those in Phase~I.
Finally, in higher fields, where only Phases IV and V are present,
the electric polarization is absent.

The left panels in Fig.~\ref{FIG:Pasummary} show the magnetic-field
dependencies for the components of $\mathbf{P}(H)$ at $T=1.1$~K. For
$\mathbf{H}\parallel\mathbf{b}$ the polarization disappears in an
abrupt way as the first-order phase transition from Phase~I to Phase
IV takes place. In contrast, for fields $\mathbf{H}\perp\mathbf{b}$
the decrease in $\mathbf{P}(H)$ is gradual, but with a clear onset
point at the saturation field. We also would like to note the
interesting behavior found in the $P_{a^{\ast}}$ component of
polarization in the magnetic field applied along $\mathbf{a^{\ast}}$
case. Here the $P_{a^{\ast}}$ component changes its sign before
fully disappearing in the paramagnetic phase. This feature is
discussed in more detail in Sec.~\ref{SEC:Phasediagrams}. An
alternative representation of the $P(H,T)$ data may be found in
Appendix~\ref{SEC:P3D}.

The actual values of saturation fields for different field
directions deserve an additional comment. Confusingly, the
saturation fields are similar for
$\mathbf{H}\parallel\mathbf{a^{\ast}},\mathbf{c^{\ast}}$ and
$\mathbf{H}\parallel\mathbf{a},\mathbf{c}$, but not for
$\mathbf{H}\parallel\mathbf{a^{\ast}},\mathbf{a}$ or
$\mathbf{H}\parallel\mathbf{c^{\ast}},\mathbf{c}$. This appears
strange, given that the mismatch between the direct and the
reciprocal space vectors is only about $13^{\circ}$. The answer to
the riddle is in the $g$ tensor, completely mapped in
Ref.~\cite{Schapers_PRB_2014_LinariteNMR}. Indeed, the principal
axis of this tensor lies between the $\mathbf{a}$ and the
$\mathbf{c}$ directions in such a way that $g_{c}\simeq g_{a}$ and
$g_{c^{\ast}}\simeq g_{a^{\ast}}$.

Finally, we would like to stress the difference between the
$\mathbf{P}(H)$ dependence observed here and the conventional
magnetoelectric
effect~\cite{Dzyaloshinskii_JETP_1959_ME,Astrov_JETP_1960_ME1,Astrov_JETP_1961_ME2}.
In a conventional magnetoelectric effect materials the polarization
change is observed in the magnetic field due to a special
symmetry-allowed bilinear term in the free energy:
$\lambda_{\alpha\beta}H_{\alpha}E_{\beta}$. In this situation
non-zero $\mathbf{H}$ results in nonzero intrinsic $\mathbf{E}$, and
$\mathbf{P}\propto \mathbf{E}$. In this case the sign of $H$ clearly
matters. In linarite the polarization is \emph{not} magnetic field
induced (It appears even in zero field!). Therefore, there is no
difference between $+H$ and $-H$ directions of the magnetic field.
Any coupling of magnetic field to polarization in our case is
related to the spin spiral structure (which is insensitive to the
sign of the magnetic field) as discussed in detail below.

\subsection{Polarization in Phase~I}
\label{SEC:ESdiscuss}

\begin{figure}
  \includegraphics[width=0.5\textwidth]{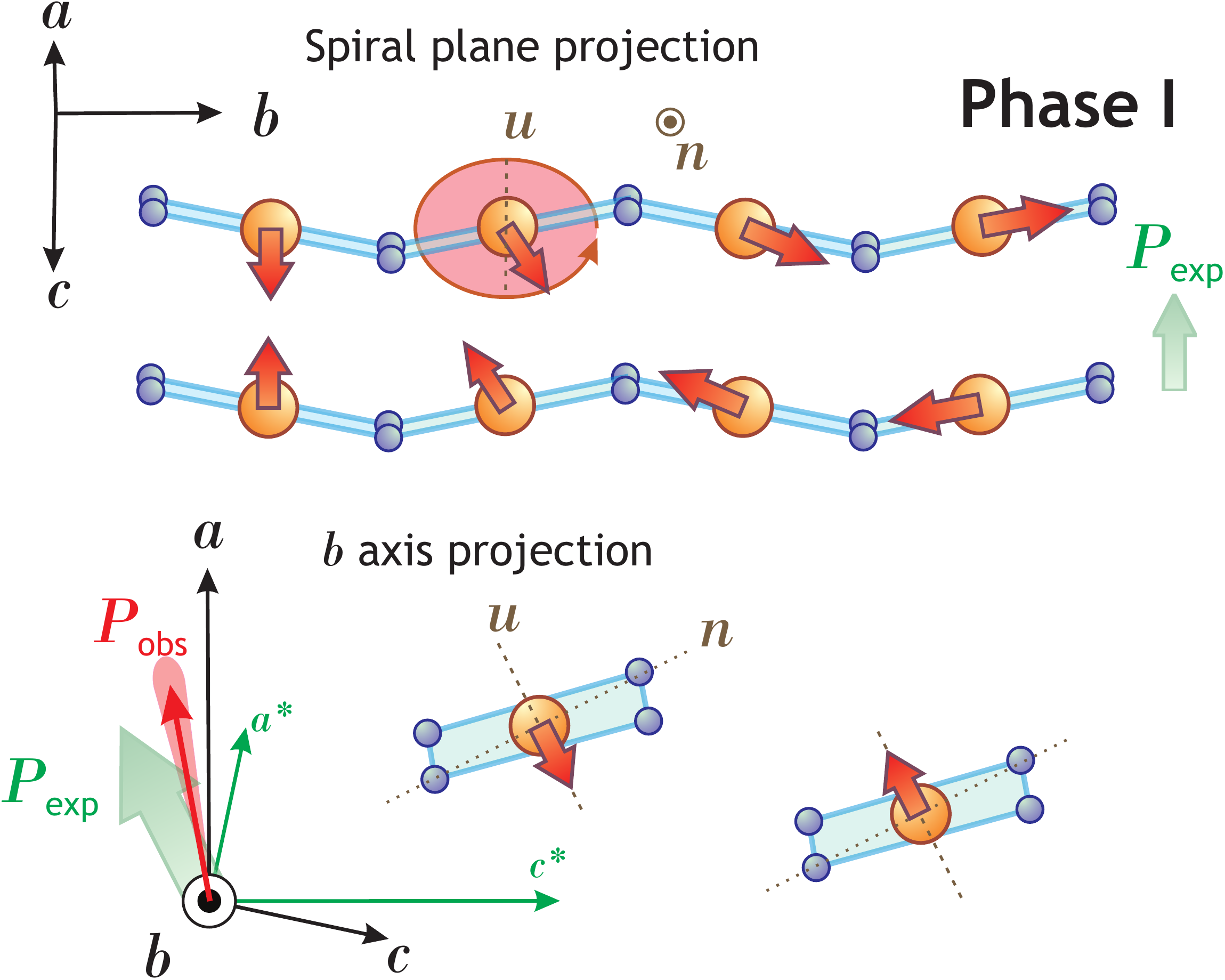}\\
  \caption{The magnetic structure of Phase~I according to
  Refs.~\cite{Willenberg_PRL_2012_LinariteFrustrated,Willenberg_PRL_2016_LinariteSDWs}.
  The direction of polarization dictated by the ``inverse
  Dzyaloshinskii--Moriya mechanism'' is indicated by the large green arrow.
  The observed polarization is shown by the red arrow, with the shaded area providing the experimental fan of
  uncertainty.
  }\label{FIG:linariteES}
\end{figure}

The magnetic structure of linarite Phase~I, found by
Willenberg~\emph{et
al.}~\cite{Willenberg_PRL_2012_LinariteFrustrated}, is an elliptic
spiral with the rotation plane significantly tilted with respect to
the crystallographic directions. It can be formally described with
the help of $\mathbf{u}$ --- a unit vector, tilted by approximately
$27^{\circ}$ off the $\mathbf{a}$ axis in the $ac$ plane. The $ub$
plane is then the plane of the spiral rotation, as shown in
Fig.~\ref{FIG:linariteES}. An additional unit vector $\mathbf{n}$,
present in this figure, is the normal vector to this rotation plane.

As follows from Fig.~\ref{FIG:Pasummary}, the polarization
components in Phase~I at zero external field are
$P_{a^{\ast}}=0.8\pm 0.1$ and $P_{c^{\ast}}=0.25\pm 0.1$~nC/cm$^{2}$
in the low-temperature limit. This is in a rough agreement with the
direction of vector $\mathbf{u}$ (see Fig.~\ref{FIG:linariteES}).
This observation is in line with the ``inverse
Dzyaloshinskii--Moryia'' or ``spin-current''
mechanism~\cite{Katsura_PRL_2005_MFmicro,Mostovoy_PRL_2006_MFmacro},
which is a typical scenario of ferroelectric polarization appearing
as a result of a spiral magnetic order. The spiral arrangement of
magnetic moments breaks the inversion symmetry \emph{within} the
spiral plane, but only if the incommensurate propagation vector
belongs to this plane as well. Mathematically it can be expressed as
$\mathbf{P}\propto[\mathbf{Q}\times\mathbf{n}]$, which, in the
present case, is exactly along the $\mathbf{u}$ direction. The exact
sign of $\mathbf{P}$ is related to the spiral's sense of rotation,
or chirality (which determines the consistent choice of signs for
$\mathbf{Q}$ and $\mathbf{n}$ vectors, with the latter defined via
the cross product of adjacent spins in the chain). A typical
situation is that in the ordered phase the spiral domains are
forming, which differ only by the sense of spiral rotation,
clockwise or counterclockwise. In centrosymmetric material one would
expect an approximately equal population of both types of domains,
which would have the opposite directions of polarization as a
consequence of the inverse Dzyaloshinskii--Moryia mechanism. An
external electric field would couple to polarization and alter the
domain population. This kind of behavior was clearly demonstrated
for LiCuVO$_4$, for
example~\cite{Mourigal_PRB_2011_LiCuVO4spincurrent}. In contrast, in
the present measurements on centrosymmetric \linarite\ the
polarization direction, and thus the spin chirality, has a preferred
direction even in the absence of biasing field. Furthermore,
field-cooling in an applied electric field is unable to reverse the
polarization. A potential explanation might be a strain in the
sample, caused by the experimental environment at low temperatures.
Improper ferroelectrics are known to be rather sensitive to elastic
perturbations~\cite{Dvorak_Ferr_1974_improperReview}. A somewhat
more exciting but speculative explanation would be an unnoticed
structural transition occurring at intermediate temperatures and
leading to the loss of inversion symmetry. One example of a symmetry
lowering transition, hardly noticeable structurally or
thermodynamically, but having a profound effect on the magnetism,
was recently discussed in
Ref.~\cite{HaelgHuvonen_PRB_2015_NTENPscaling}.

Rigorously speaking, this unexpected symmetry breaking in linarite
remains enigmatic and further work is needed to clarify this issue.

\subsection{Polarization in Phase III}

\begin{figure}
  \includegraphics[width=0.5\textwidth]{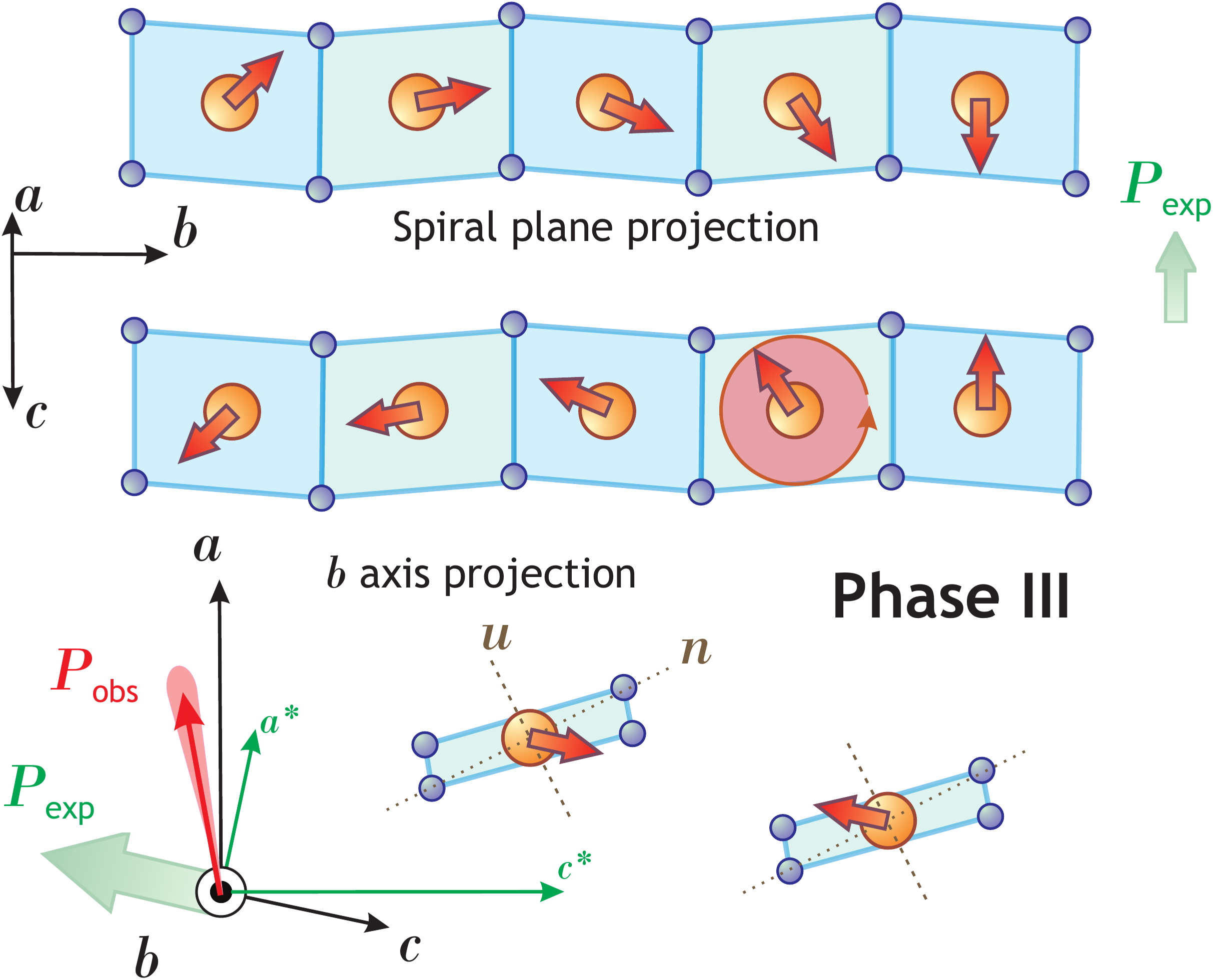}\\
  \caption{Same as Fig.~\ref{FIG:linariteES}, but for the
  field-induced Phase III. The observed direction of polarization is inconsistent with that deduced from the proposed magnetic structure.
  }\label{FIG:linariteCS}
\end{figure}

Our measurements show that Phase~I is not the only multiferroic
phase of linarite. We also find clear signatures of ferroelectricity
(mainly visible in temperature scans) in Phase III. Neutron
diffraction studies in Ref.~\cite{Willenberg_PRL_2016_LinariteSDWs}
have suggested this phase to be of a mixed nature, with the
intensity from incommensurate magnetic Bragg peaks gradually
shifting to commensurate magnetic Bragg peaks (belonging to Phase
IV). At the same time, even though the incommensurate propagation
vector in Phase III is the same as that in Phase~I,
$\mathbf{Q}=(0,\text{ }-0.186,\text{ }0.5)$~r.l.u., the spin
arrangement is claimed to be different. It  was identified as a
circular helix with the rotation plane matching the $(\mathbf{b,c})$
plane of the crystal. As the structure is still a spin spiral, the
presence of electric polarization is not surprising. However, the
observed direction of $\mathbf{P}$ is \emph{inconsistent} with the
proposed spin arrangement, as illustrated in
Fig.~\ref{FIG:linariteCS}. Indeed, in the inverse
Dzyaloshinskii--Moriya mechanism necessarily
$\mathbf{P}\propto[\mathbf{Q}\times\mathbf{n}]$. Therefore, a
$(\mathbf{b,c})$ planar spiral strictly requires the polarization to
lie \emph{within} the spiral plane. In contrast, experimentally the
largest polarization component is observed along direction
$\mathbf{a^{\ast}}$, which is \emph{normal} to this plane.

There are known cases of multiferroics with a relation between the
spin structure and the polarization vector that is much more complex
than suggested by the straightforward inverse Dzyaloshinskii--Moryia
mechanism~\cite{Katsura_PRL_2005_MFmicro,Mostovoy_PRL_2006_MFmacro}.
A very representative example is the triangular lattice
antiferromagnet
RbFe(MoO$_4$)$_2$~\cite{Kenzelmann_PRL_2007_TriangularFerro}, where
the electric polarization is exactly orthogonal to the helimagnetic
planes. Cases like this require a more advanced symmetry-based
treatment, as described in detail by
Harris~\cite{Harris_PRB_2007_MFlandau}. In linarite, however, the
proposed model of the I--III phase transition does not seem to
involve a significant change in the magnetic-state symmetry. Hence,
it is not clear why it would invoke a completely different type of
coupling between magnetism and polarization. This is especially
strange given that the observed direction of $\mathbf{P}$ remains
unchanged across the transition. Trying to make sense of this, we
note that magnets with complex interactions often have very peculiar
spin structures with multiple propagation vectors. One example is
the spiral antiferromagnet Ba$_2$CuGe$_2$O$_7$, in which a very
special antiferromagnetic cone phase described by simultaneous
commensurate and incommensurate wave vectors was
found~\cite{Muehlbauer_PRB_2011_BaCuGeOmultiK}. The phenomenology of
neutron diffraction observations in Ba$_2$CuGe$_2$O$_7$ is
remarkably similar to that in Phase III in
linarite~\cite{Willenberg_PRL_2016_LinariteSDWs}. This analogy hints
that Phase III in linarite may, in fact, also be a two-$Q$
structure, rather than a mixed phase like Phase II. This idea is
consistent with the stable character of electric anomalies in Phase
III, contrasting with the history-dependent behavior found in Phase
II. Some more details regarding this possible structure are given in
Appendix~\ref{SEC:multiQ}. To summarize this discussion we just
briefly note that in the case of Ba$_2$CuGe$_2$O$_7$ the key to
stabilizing this rather peculiar phase was the interplay of external
field and anisotropic interactions (especially antisymmetric
Dzyaloshinskii--Moriya
interactions~\cite{Dzyaloshinskii_JChemPhysSol_1958_DM,Moriya_PR_1960_DM}).
In linarite the dramatic anisotropy of the phase diagram makes
apparent the presence of non-negligible anisotropy in the
interactions. At the same time, the antisymmetric geometry of the
superexchange bond between the nearest Cu$^{2+}$ neighbors allows
the presence of Dzyaloshinskii--Moriya interactions, staggered along
the chain.

\subsection{Multiferroic metastability at low temperatures}

As shown in Fig.~\ref{FIG:IaHb_HYST}, the situation with the
isothermal magnetoelectric current becomes quite complicated as one
reaches the region denoted II in the phase diagram
(Fig.~\ref{FIG:linaritePhD}). The evidence from the previous studies
is that this metastable region corresponds to the coexistence of
domains of Phases I  and IV. This picture agrees well with our
present observations. In this regime a well-defined anomaly
corresponding to a first-order transition is replaced by a family of
extremely sharp peaks that show a strong history dependence. In
fact, each such spike manifests the loss of stability of a single
Phase~I domain (with increasing field) or Phase IV domain (on
decreasing field). The observed spikes are just a differential
multiferroic analog of a familiar Barkhausen effect in ferromagnets.

\subsection{Magnetic phase diagrams from electric measurements}
\label{SEC:Phasediagrams}

\begin{figure}
  \includegraphics[width=0.5\textwidth]{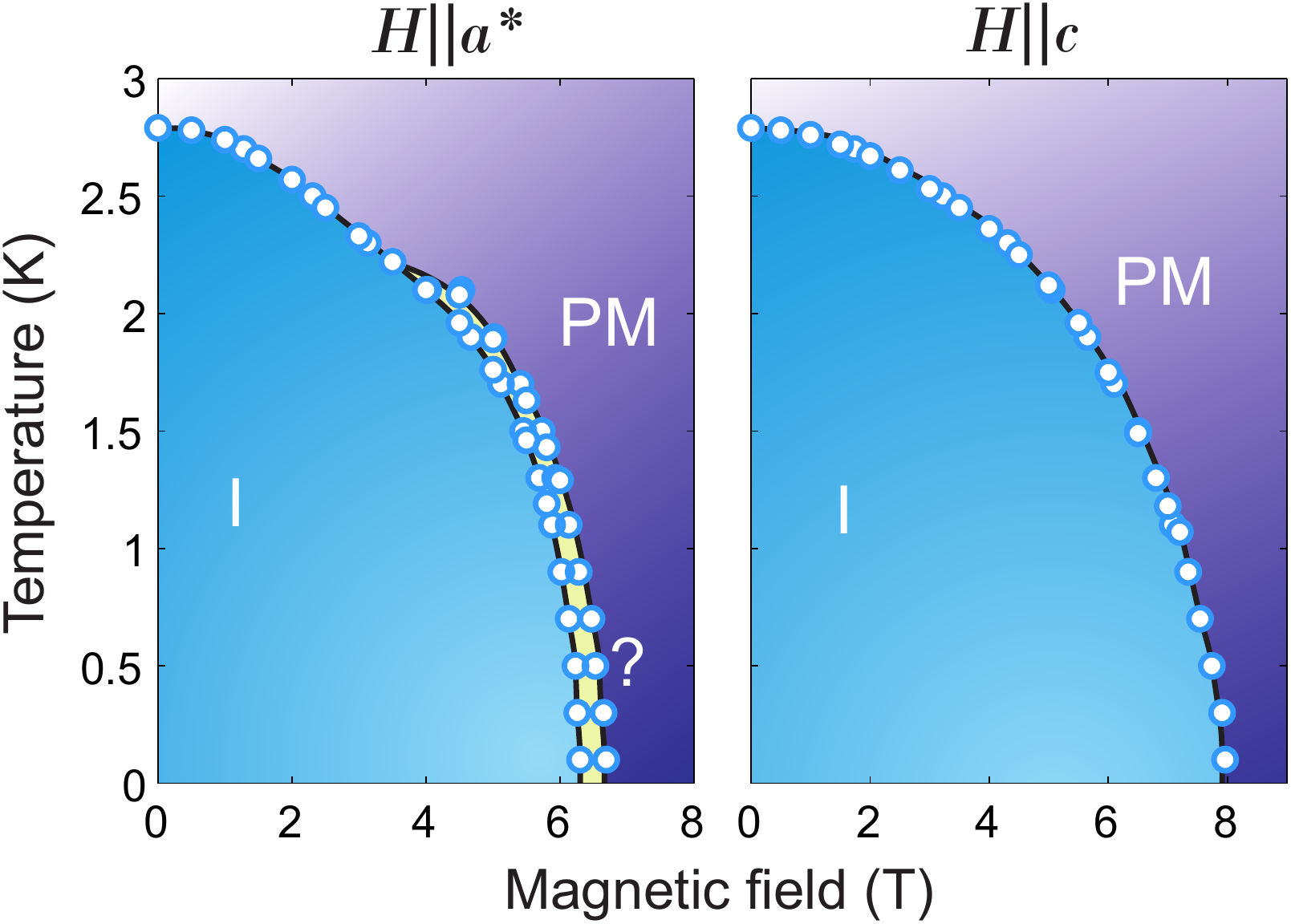}\\
  \caption{Magnetic phase diagram of linarite deduced from electric current measurements for magnetic fields applied
  transverse to the $\mathbf{b}$ direction. Left: Field along $\mathbf{a}^{\ast}$. The two
  dominant phases are Phase~I and the disordered paramagnetic state. In high fields there is an additional intervening region,
  although it is not clear whether it represents a distinct thermodynamic phase.
  Right: Field along the $\mathbf{c}$ direction. Symbols are experimental
  data;
  solid lines are guides for the eye.}\label{FIG:linaritePhDtrans}
\end{figure}

The collected current data conveniently provide us with a way to
reconstruct the magnetic phase diagram. The summary of our
measurements is shown in Figs.~\ref{FIG:linaritePhD} and
\ref{FIG:linaritePhDtrans}. The phase diagram for the field applied
in the $\mathbf{b}$ direction is clearly the most interesting. Even
though we find no electric activity in Phases IV and V, all the
other ordered phases are proved to be multiferroic in nature.

We also find a peculiar behavior of the polarization vector
(polarization reversal) near saturation for a magnetic field applied
in the $\mathbf{a^{\ast}}$ direction. According to Sch\"apers
\emph{et al}.~\cite{Schapers_PRB_2013_LinariteBulk}, this  phase
diagram should contain a single elliptical spiral
phase~\footnote{Nonetheless, we note that the last non-featureless
specific heat curve in the corresponding panel of Fig.~9 in
Ref.~\cite{Schapers_PRB_2013_LinariteBulk} may give some room for
speculation.}. The abrupt reversal of polarization observed in our
experiments is not necessarily indicative of a thermodynamic phase
transition. It may just be the result of strong deformation of the
spin structure by the magnetic field, as it becomes almost
polarized. Nor can one fully exclude a spurious origin of this
feature. The presence of misaligned grain in the sample may in
principle result in this kind of behavior. On the other hand, the
data for $\mathbf{H}\parallel \mathbf{b}$ and $\mathbf{H}\parallel
\mathbf{c}$ taken from \emph{the same sample} show no phase boundary
``splitting'' in applied magnetic fields. Whether or not there is an
additional phase in this geometry near saturation remains an open
question.

The phase diagrams for magnetic fields applied in the $\mathbf{a}$,
$\mathbf{c}$ and $\mathbf{c}^{\ast}$ directions undoubtedly contain
just a single ordered phase with a conventionally looking phase
boundary. In Fig.~\ref{FIG:linaritePhDtrans} we show just one such
case with $\mathbf{H}\parallel\mathbf{c}$.

\section{Summary}
At least three of linarite's magnetic phases support spontaneous
electric polarization: (i) The principal spin spiral state (Phase~I)
appears to be a classic ``reverse Dzyaloshinskii--Moriya'' improper
ferroelectric. The observed direction of polarization is fully
consistent with the magnetic structure proposed in
Refs.~\cite{Willenberg_PRL_2012_LinariteFrustrated,Willenberg_PRL_2016_LinariteSDWs}.
(ii) Also in agreement with
Ref.~\cite{Schapers_PRB_2013_LinariteBulk}, region II, found between
the spiral Phase~I and the collinear Phase IV, is actually a
phase-separation regime, showing a typical history-dependent
behavior of polarization. (iii) The polarization observed in Phase
III is \emph{not} consistent with the phase-separated magnetic state
proposed in Ref.~\cite{Willenberg_PRL_2016_LinariteSDWs}. Rather, it
appears to be a stable thermodynamic phase and may be a complex
multi-$Q$ spin structure. (iv) A new region of polarization
reversal, which may or may not be a distinct thermodynamic phase, is
identified close to saturation for a magnetic field applied along
the $\mathbf{a}^{\ast}$ direction.

\acknowledgments

This work was supported by the Swiss National Science Foundation,
Division 2. We also thank Dr. S. Gvasaliya (ETH Z\"{u}rich) for
support with experiments.

\appendix

\section{The polarization map}
\label{SEC:P3D}

\begin{figure}
  \includegraphics[width=0.5\textwidth]{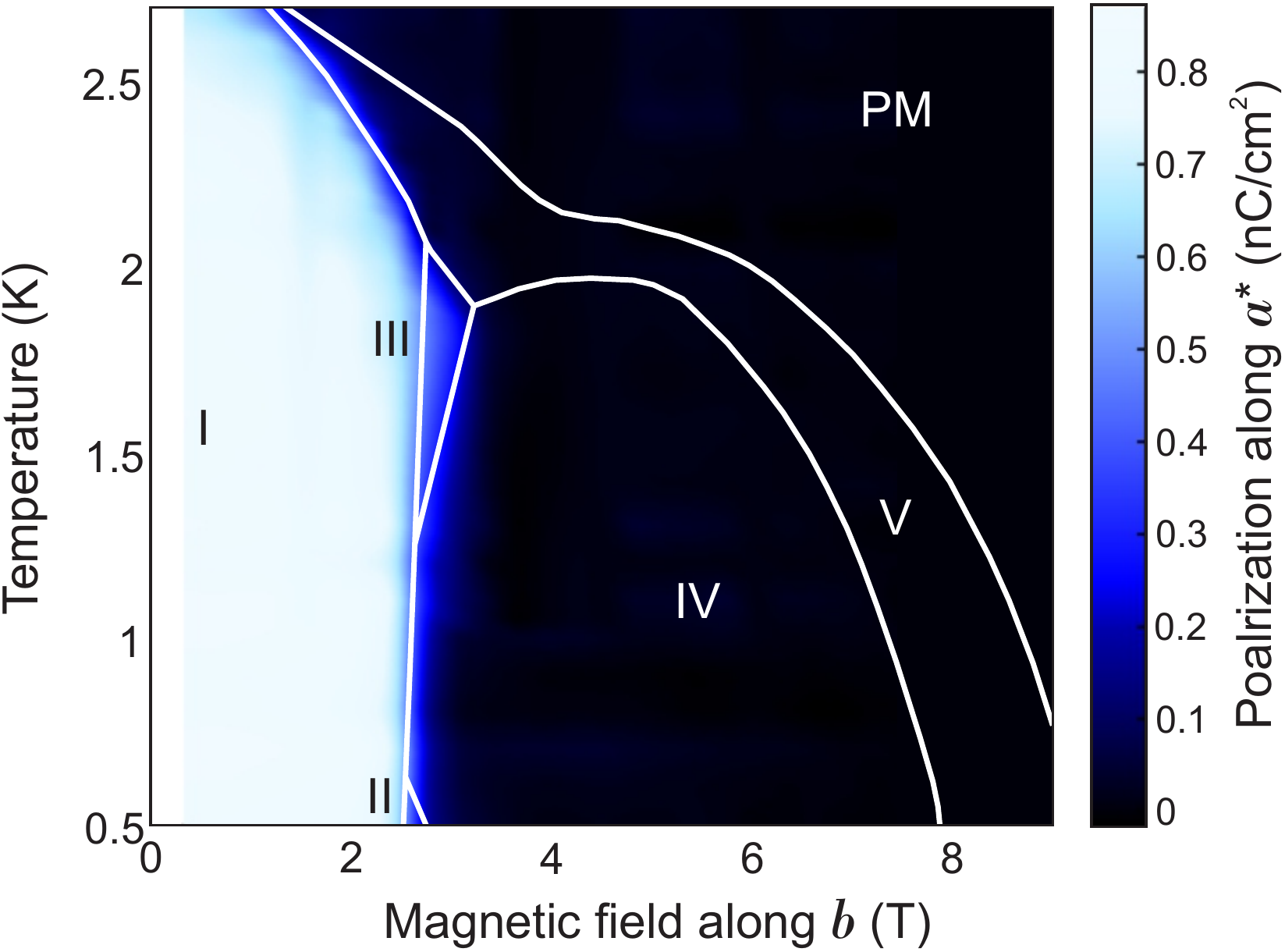}\\
  \caption{False color map of $P_{a^{\ast}}$ in a magnetic field along the $b$ direction. Solid lines correspond to the phase boundaries
  according to~\cite{Willenberg_PRL_2016_LinariteSDWs}.}\label{FIG:P3D}
\end{figure}

The number of collected data allows us to reconstruct the $P(H,T)$
surface. An example is shown in Fig.~\ref{FIG:P3D}. Here the
$P_{a^{\ast}}$ component, reconstructed from the field scans is
plotted. The region very close to $H=0$ is omitted here (due to
nonlinearity of the sweep rate at the very beginning of the scan).
We also show the data at temperatures above Phase II only, as within
this metastable phase no equilibrium polarization value can be
consistently defined.

\section{The multi-$Q$ structure}
\label{SEC:multiQ}
\subsection{A toy model}

\begin{figure}
  \includegraphics[width=0.5\textwidth]{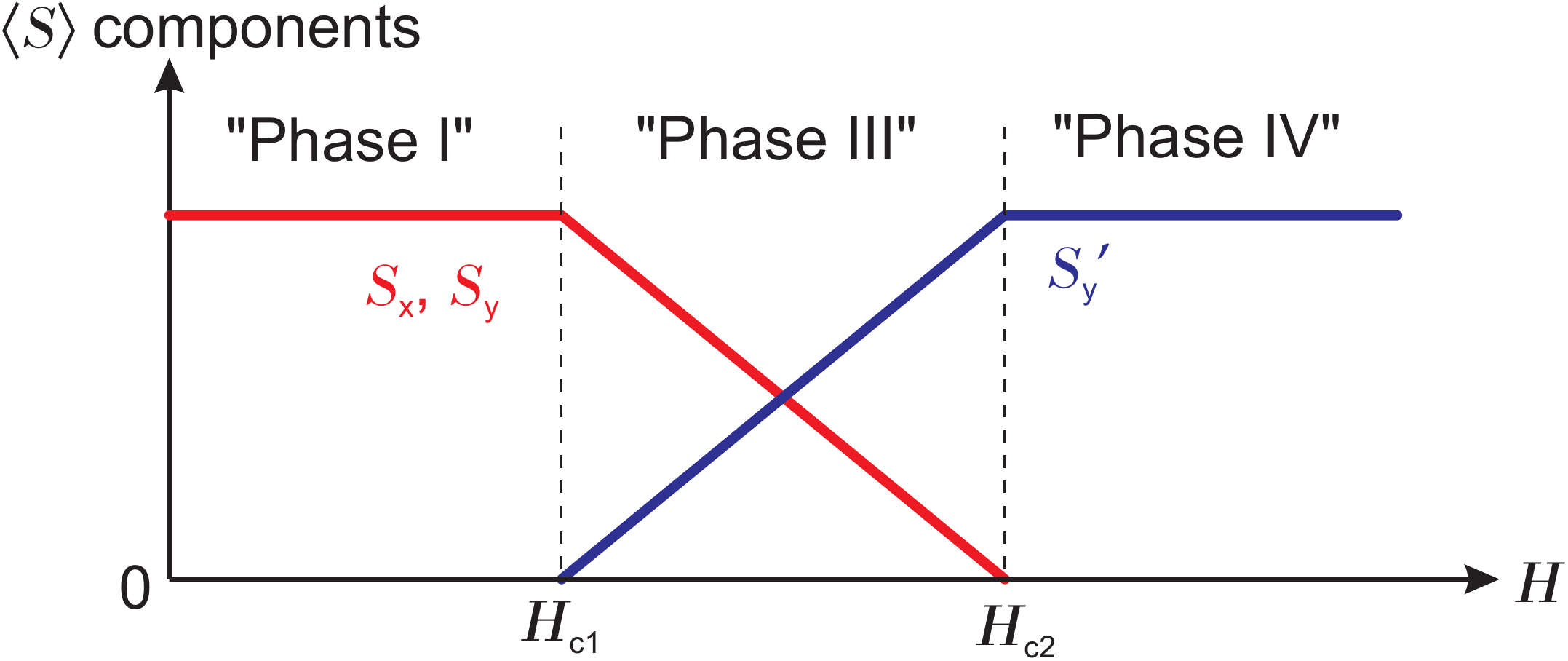}\\
  \caption{Sketch of magnetic field dependence of the spin texture components $S_x$, $S_y$ (of ``Phase~I'') and $S'_y$ (of ``Phase IV'') in the toy model.
  There is a region between $H_{c1}$ and $H_{c2}$ where both types of order coexist [as shown by Eq.~(\ref{EQ:multiQtoyfull})]. This region corresponds
  to ``Phase III''.}\label{FIG:multiQ_spins}
\end{figure}

\begin{figure}
  \includegraphics[width=0.5\textwidth]{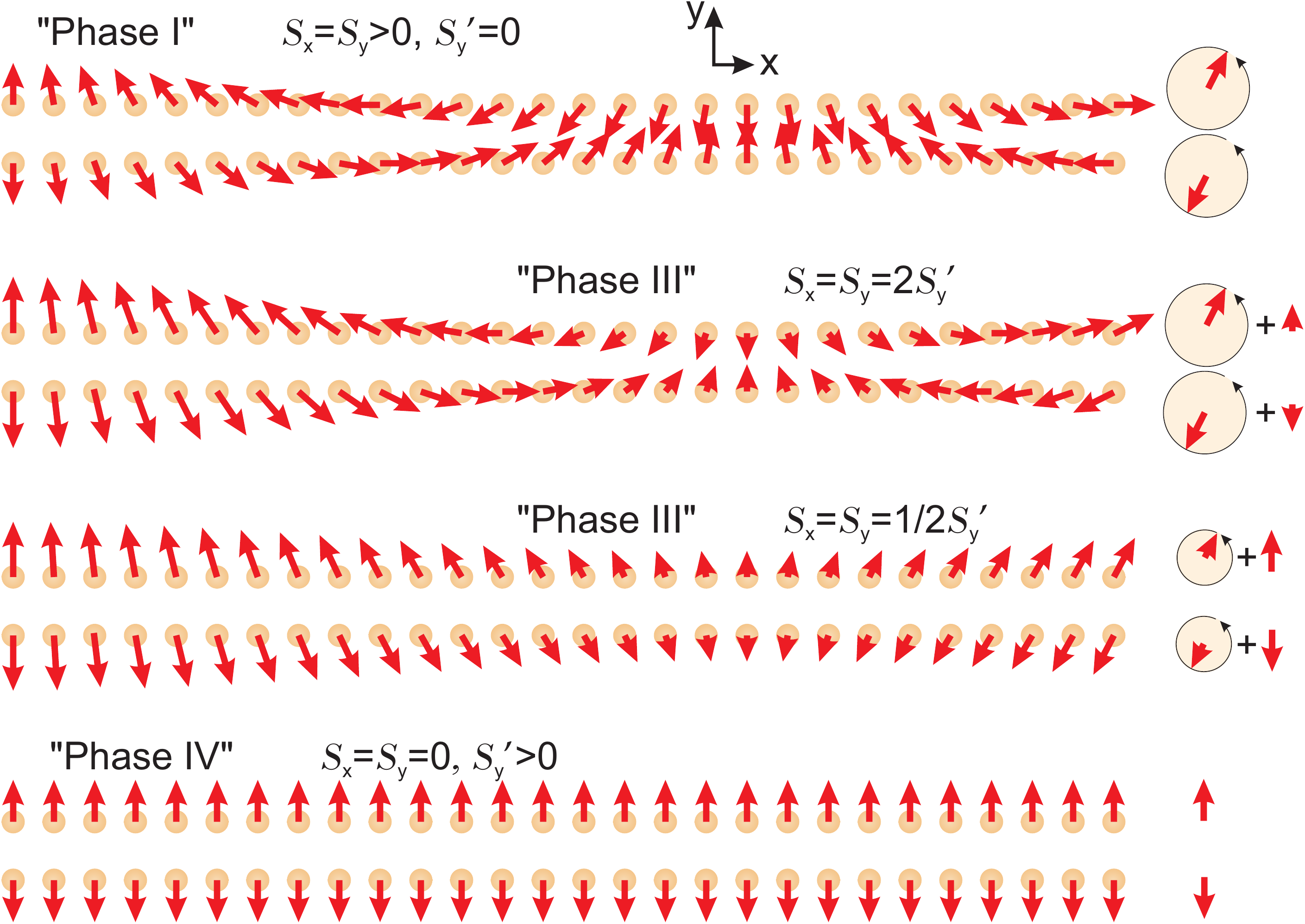}\\
  \caption{Spin textures of the toy model, corresponding to different phases (and different magnetic-field strength).
  First comes the simple spiral ``Phase~I'' at $H<H_{c1}$. Then there is ``Phase III'' [as given by Eq.~(\ref{EQ:multiQtoyfull})], shown in two
  regimes: with a dominating spiral component ($H\gtrsim H_{c1}$) and with a dominating antiferromagnetic component ($H\lesssim H_{c2}$).
  The last phase is the straight antiferromagnetic ``Phase IV'' at $H>H_{c2}$.}\label{FIG:multiQ}
\end{figure}

Before describing the plausible multi-$Q$ structure of Phase III in
linarite, interpolating between Phases I and IV, we would like to
consider a simplified toy model. This model serves as a nice
illustration of the crossover between incommensurate and
commensurate structures and is free of two major complications
present in linarite: the ellipticity of the spiral phase and the
noncoplanarity of the spin structures between Phases I and IV. In
this two-dimensional toy model the low-field ``Phase~I'' would be
simply a circular spiral in the $\mathbf{xy}$ plane ($\mathbf{x}$ is
the direction along the chain; the analog of $\mathbf{b}$ in the
real linarite structure, and $\mathbf{y}$ is the analog of
$\mathbf{c}$). Then the ``Phase~I'' order is described as
$\aver{\mathbf{S}_{nm}}=\mathbf{S}_{y}\cos(2\pi\mathbf{Q}_{I}\mathbf{r}_{nm})+\mathbf{S}_{x}\sin(2\pi\mathbf{Q}_{I}\mathbf{r}_{nm})$,
with $|\mathbf{S}_{x}|=|\mathbf{S}_{y}|$ and
$\mathbf{Q}_{I}=(\epsilon, 0.5)$ (here $\epsilon$ is the
incommensuration parameter; in Fig.~\ref{FIG:multiQ} $\epsilon=1/36$
is taken). The vector $\mathbf{r}_{mn}=n\mathbf{x}+m\mathbf{y}$
simply enumerates the spin sites ($n$ along the chain, $m$ between
the chains). In contrast, the commensurate order in ``Phase IV'' is
simply
$\aver{\mathbf{S}_{nm}}=\mathbf{S'}_{y}\cos(2\pi\mathbf{Q}_{IV}\mathbf{r}_{nm})$,
where $\mathbf{Q}_{IV}=(0, 0.5)$.

Then the essential idea of constructing the spin texture of ``Phase
III'' is to consider a linear combination of the two structures
described above. The description of this complex ordering is given
as:

\begin{equation}\label{EQ:multiQtoyfull}
\begin{aligned}
    \aver{\mathbf{S}_{nm}}=\mathbf{S}_{y}\cos(2\pi\mathbf{Q}_{I}\mathbf{r}_{nm})+\mathbf{S}_{x}\sin(2\pi\mathbf{Q}_{I}\mathbf{r}_{nm})\\
    +\mathbf{S'}_{y}cos(2\pi\mathbf{Q}_{IV}\mathbf{r}_{nm}),\\
\end{aligned}
\end{equation}

Within the toy model we may assume the dependence of
$\mathbf{S}_{\mathbf{x},\mathbf{y}}$ and $\mathbf{S}'_{\mathbf{y}}$
on the external magnetic field as described in
Fig.~\ref{FIG:multiQ_spins}. Then in the region $H_{c1}<H<H_{c2}$ we
find a complex type of order exhibiting propagation vectors
$\mathbf{Q}_{I}$ and $\mathbf{Q}_{IV}$ simultaneously. The examples
of the resulting spin structure in different regimes are given in
Fig.~\ref{FIG:multiQ}.

\subsection{Application to the actual Phase III of linarite}

The structure of the actual Phase III of linarite is constructed
from the known structures of Phases I and IV in direct analogy with
the toy model, described above. The plausible spin arrangement is
described as

\begin{equation}\label{EQ:multiQiii}
\begin{aligned}
    \aver{\mathbf{S}(\mathbf{r})}=\mathbf{S}_{\mathbf{u}}\cos(2\pi\mathbf{Q}_{I}\mathbf{r})+\mathbf{S}_{\mathbf{b}}\sin(2\pi\mathbf{Q}_{I}\mathbf{r})\\
    +\mathbf{S}_{\mathbf{c}}cos(2\pi\mathbf{Q}_{IV}\mathbf{r}).\\
\end{aligned}
\end{equation}

Again, $\mathbf{Q}_{I}=(0,0.186,0.5)$ and
$\mathbf{Q}_{IV}=(0,0,0.5)$ are the ``original'' propagation vectors
of limiting single-$Q$ structures. Vector $\mathbf{u}$ is the same
as described in Sec.~\ref{SEC:ESdiscuss}. However, the complication
[which makes the visualization of Eq.~(\ref{EQ:multiQiii}) not very
useful] is that the spin vector $\mathbf{S}_{\mathbf{c}}$ does not
belong to the plane given by vectors
$\mathbf{S}_{\mathbf{u}}-\mathbf{S}_{\mathbf{b}}$ and is not
orthogonal to this plane either. Ellipticity of the spiral
contribution ($\mathbf{S}_{\mathbf{b}}>\mathbf{S}_{\mathbf{u}}$)
also adds to the overall complexity of the resulting structure. One
has to keep in mind, that there is a significant ferromagnetic
moment along the field direction $\mathbf{b}$ present in Phase III
as well (as the material is approximately one-third magnetized in
this phase). This contribution is not included in
Eq.~(\ref{EQ:multiQiii}).

\bibliography{The_Library}

\begin{thebibliography}{34}%
\makeatletter
\providecommand \@ifxundefined [1]{%
 \@ifx{#1\undefined}
}%
\providecommand \@ifnum [1]{%
 \ifnum #1\expandafter \@firstoftwo
 \else \expandafter \@secondoftwo
 \fi
}%
\providecommand \@ifx [1]{%
 \ifx #1\expandafter \@firstoftwo
 \else \expandafter \@secondoftwo
 \fi
}%
\providecommand \natexlab [1]{#1}%
\providecommand \enquote  [1]{``#1''}%
\providecommand \bibnamefont  [1]{#1}%
\providecommand \bibfnamefont [1]{#1}%
\providecommand \citenamefont [1]{#1}%
\providecommand \href@noop [0]{\@secondoftwo}%
\providecommand \href [0]{\begingroup \@sanitize@url \@href}%
\providecommand \@href[1]{\@@startlink{#1}\@@href}%
\providecommand \@@href[1]{\endgroup#1\@@endlink}%
\providecommand \@sanitize@url [0]{\catcode `\\12\catcode `\$12\catcode
  `\&12\catcode `\#12\catcode `\^12\catcode `\_12\catcode `\%12\relax}%
\providecommand \@@startlink[1]{}%
\providecommand \@@endlink[0]{}%
\providecommand \url  [0]{\begingroup\@sanitize@url \@url }%
\providecommand \@url [1]{\endgroup\@href {#1}{\urlprefix }}%
\providecommand \urlprefix  [0]{URL }%
\providecommand \Eprint [0]{\href }%
\providecommand \doibase [0]{http://dx.doi.org/}%
\providecommand \selectlanguage [0]{\@gobble}%
\providecommand \bibinfo  [0]{\@secondoftwo}%
\providecommand \bibfield  [0]{\@secondoftwo}%
\providecommand \translation [1]{[#1]}%
\providecommand \BibitemOpen [0]{}%
\providecommand \bibitemStop [0]{}%
\providecommand \bibitemNoStop [0]{.\EOS\space}%
\providecommand \EOS [0]{\spacefactor3000\relax}%
\providecommand \BibitemShut  [1]{\csname bibitem#1\endcsname}%
\let\auto@bib@innerbib\@empty
\bibitem [{\citenamefont {Khomskii}(2006)}]{Khomskii_JMMM_2006_MFreview}%
  \BibitemOpen
  \bibfield  {author} {\bibinfo {author} {\bibfnamefont {D.~I.}\ \bibnamefont
  {Khomskii}},\ }\bibfield  {title} {\enquote {\bibinfo {title} {Multiferroics:
  Different ways to combine magnetism and ferroelectricity},}\ }\href
  {http://www.sciencedirect.com/science/article/pii/S0304885306004239}
  {\bibfield  {journal} {\bibinfo  {journal} {J. Magn. Magn. Mater.}\ }\textbf
  {\bibinfo {volume} {306}},\ \bibinfo {pages} {1} (\bibinfo {year}
  {2006})}\BibitemShut {NoStop}%
\bibitem [{\citenamefont {Cheong}\ and\ \citenamefont
  {Mostovoy}(2007)}]{CheongMosotovoy_NatMat_2007_NatureReview}%
  \BibitemOpen
  \bibfield  {author} {\bibinfo {author} {\bibfnamefont {S.-W.}\ \bibnamefont
  {Cheong}}\ and\ \bibinfo {author} {\bibfnamefont {M.}~\bibnamefont
  {Mostovoy}},\ }\bibfield  {title} {\enquote {\bibinfo {title}
  {{Multiferroics: a magnetic twist for ferroelectricity}},}\ }\href
  {http://www.nature.com/nmat/journal/v6/n1/abs/nmat1804.html} {\bibfield
  {journal} {\bibinfo  {journal} {Nat. Mater.}\ }\textbf {\bibinfo {volume}
  {6}},\ \bibinfo {pages} {13} (\bibinfo {year} {2007})}\BibitemShut {NoStop}%
\bibitem [{\citenamefont {Tokura}\ \emph {et~al.}(2014)\citenamefont {Tokura},
  \citenamefont {Seki},\ and\ \citenamefont
  {Nagaosa}}]{Tokura_RevProgPhys_2014_MFreview}%
  \BibitemOpen
  \bibfield  {author} {\bibinfo {author} {\bibfnamefont {Y.}~\bibnamefont
  {Tokura}}, \bibinfo {author} {\bibfnamefont {S.}~\bibnamefont {Seki}}, \ and\
  \bibinfo {author} {\bibfnamefont {N.}~\bibnamefont {Nagaosa}},\ }\bibfield
  {title} {\enquote {\bibinfo {title} {Multiferroics of spin origin},}\ }\href
  {http://iopscience.iop.org/0034-4885/77/7/076501} {\bibfield  {journal}
  {\bibinfo  {journal} {Rep. Prog. Phys.}\ }\textbf {\bibinfo {volume} {77}},\
  \bibinfo {pages} {076501} (\bibinfo {year} {2014})}\BibitemShut {NoStop}%
\bibitem [{\citenamefont {Goto}\ \emph {et~al.}(2004)\citenamefont {Goto},
  \citenamefont {Kimura}, \citenamefont {Lawes}, \citenamefont {Ramirez},\ and\
  \citenamefont {Tokura}}]{GotoKimura_PRL_2004_GiantMCeffect}%
  \BibitemOpen
  \bibfield  {author} {\bibinfo {author} {\bibfnamefont {T.}~\bibnamefont
  {Goto}}, \bibinfo {author} {\bibfnamefont {T.}~\bibnamefont {Kimura}},
  \bibinfo {author} {\bibfnamefont {G.}~\bibnamefont {Lawes}}, \bibinfo
  {author} {\bibfnamefont {A.~P.}\ \bibnamefont {Ramirez}}, \ and\ \bibinfo
  {author} {\bibfnamefont {Y.}~\bibnamefont {Tokura}},\ }\bibfield  {title}
  {\enquote {\bibinfo {title} {{Ferroelectricity and Giant Magnetocapacitance
  in Perovskite Rare-Earth Manganites}},}\ }\href {\doibase
  10.1103/PhysRevLett.92.257201} {\bibfield  {journal} {\bibinfo  {journal}
  {Phys. Rev. Lett.}\ }\textbf {\bibinfo {volume} {92}},\ \bibinfo {pages}
  {257201} (\bibinfo {year} {2004})}\BibitemShut {NoStop}%
\bibitem [{\citenamefont {Kagawa}\ \emph {et~al.}(2009)\citenamefont {Kagawa},
  \citenamefont {Mochizuki}, \citenamefont {Onose}, \citenamefont {Murakawa},
  \citenamefont {Kaneko}, \citenamefont {Furukawa},\ and\ \citenamefont
  {Tokura}}]{KagawaMochizuki_PRL_2009_GiantMCdomains}%
  \BibitemOpen
  \bibfield  {author} {\bibinfo {author} {\bibfnamefont {F.}~\bibnamefont
  {Kagawa}}, \bibinfo {author} {\bibfnamefont {M.}~\bibnamefont {Mochizuki}},
  \bibinfo {author} {\bibfnamefont {Y.}~\bibnamefont {Onose}}, \bibinfo
  {author} {\bibfnamefont {H.}~\bibnamefont {Murakawa}}, \bibinfo {author}
  {\bibfnamefont {Y.}~\bibnamefont {Kaneko}}, \bibinfo {author} {\bibfnamefont
  {N.}~\bibnamefont {Furukawa}}, \ and\ \bibinfo {author} {\bibfnamefont
  {Y.}~\bibnamefont {Tokura}},\ }\bibfield  {title} {\enquote {\bibinfo {title}
  {{Dynamics of Multiferroic Domain Wall in Spin-Cycloidal Ferroelectric
  ${\mathrm{DyMnO}}_{3}$}},}\ }\href {\doibase 10.1103/PhysRevLett.102.057604}
  {\bibfield  {journal} {\bibinfo  {journal} {Phys. Rev. Lett.}\ }\textbf
  {\bibinfo {volume} {102}},\ \bibinfo {pages} {057604} (\bibinfo {year}
  {2009})}\BibitemShut {NoStop}%
\bibitem [{\citenamefont {Schrettle}\ \emph {et~al.}(2013)\citenamefont
  {Schrettle}, \citenamefont {Krohns}, \citenamefont {Lunkenheimer},
  \citenamefont {Loidl}, \citenamefont {Wulf}, \citenamefont {Yankova},\ and\
  \citenamefont {Zheludev}}]{Schrettle_PRB_2013_SulMF}%
  \BibitemOpen
  \bibfield  {author} {\bibinfo {author} {\bibfnamefont {F.}~\bibnamefont
  {Schrettle}}, \bibinfo {author} {\bibfnamefont {S.}~\bibnamefont {Krohns}},
  \bibinfo {author} {\bibfnamefont {P.}~\bibnamefont {Lunkenheimer}}, \bibinfo
  {author} {\bibfnamefont {A.}~\bibnamefont {Loidl}}, \bibinfo {author}
  {\bibfnamefont {E.}~\bibnamefont {Wulf}}, \bibinfo {author} {\bibfnamefont
  {T.}~\bibnamefont {Yankova}}, \ and\ \bibinfo {author} {\bibfnamefont
  {A.}~\bibnamefont {Zheludev}},\ }\bibfield  {title} {\enquote {\bibinfo
  {title} {Magnetic-field induced multiferroicity in a quantum critical
  frustrated spin liquid},}\ }\href {\doibase 10.1103/PhysRevB.87.121105}
  {\bibfield  {journal} {\bibinfo  {journal} {Phys. Rev. B}\ }\textbf {\bibinfo
  {volume} {87}},\ \bibinfo {pages} {121105} (\bibinfo {year}
  {2013})}\BibitemShut {NoStop}%
\bibitem [{\citenamefont {Povarov}\ \emph {et~al.}(2015)\citenamefont
  {Povarov}, \citenamefont {Reichert}, \citenamefont {Wulf},\ and\
  \citenamefont {Zheludev}}]{PovarovReichert_PRB_2015_Sulf}%
  \BibitemOpen
  \bibfield  {author} {\bibinfo {author} {\bibfnamefont {{\relax K.
  Yu}}~\bibnamefont {Povarov}}, \bibinfo {author} {\bibfnamefont
  {A.}~\bibnamefont {Reichert}}, \bibinfo {author} {\bibfnamefont
  {E.}~\bibnamefont {Wulf}}, \ and\ \bibinfo {author} {\bibfnamefont
  {A.}~\bibnamefont {Zheludev}},\ }\bibfield  {title} {\enquote {\bibinfo
  {title} {{Giant dielectric nonlinearities at a magnetic Bose-Einstein
  condensation}},}\ }\href {\doibase 10.1103/PhysRevB.92.140410} {\bibfield
  {journal} {\bibinfo  {journal} {Phys. Rev. B}\ }\textbf {\bibinfo {volume}
  {92}},\ \bibinfo {pages} {140410} (\bibinfo {year} {2015})}\BibitemShut
  {NoStop}%
\bibitem [{\citenamefont {Park}\ \emph {et~al.}(2007)\citenamefont {Park},
  \citenamefont {Choi}, \citenamefont {Zhang},\ and\ \citenamefont
  {Cheong}}]{Park_PRL_2007_LiCu2O2Multiferr}%
  \BibitemOpen
  \bibfield  {author} {\bibinfo {author} {\bibfnamefont {S.}~\bibnamefont
  {Park}}, \bibinfo {author} {\bibfnamefont {Y.~J.}\ \bibnamefont {Choi}},
  \bibinfo {author} {\bibfnamefont {C.~L.}\ \bibnamefont {Zhang}}, \ and\
  \bibinfo {author} {\bibfnamefont {S-W.}\ \bibnamefont {Cheong}},\ }\bibfield
  {title} {\enquote {\bibinfo {title} {Ferroelectricity in an $s=1/2$ chain
  cuprate},}\ }\href {\doibase 10.1103/PhysRevLett.98.057601} {\bibfield
  {journal} {\bibinfo  {journal} {Phys. Rev. Lett.}\ }\textbf {\bibinfo
  {volume} {98}},\ \bibinfo {pages} {057601} (\bibinfo {year}
  {2007})}\BibitemShut {NoStop}%
\bibitem [{\citenamefont {Mourigal}\ \emph {et~al.}(2011)\citenamefont
  {Mourigal}, \citenamefont {Enderle}, \citenamefont {Kremer}, \citenamefont
  {Law},\ and\ \citenamefont {F\aa{}k}}]{Mourigal_PRB_2011_LiCuVO4spincurrent}%
  \BibitemOpen
  \bibfield  {author} {\bibinfo {author} {\bibfnamefont {M.}~\bibnamefont
  {Mourigal}}, \bibinfo {author} {\bibfnamefont {M.}~\bibnamefont {Enderle}},
  \bibinfo {author} {\bibfnamefont {R.~K.}\ \bibnamefont {Kremer}}, \bibinfo
  {author} {\bibfnamefont {J.~M.}\ \bibnamefont {Law}}, \ and\ \bibinfo
  {author} {\bibfnamefont {B.}~\bibnamefont {F\aa{}k}},\ }\bibfield  {title}
  {\enquote {\bibinfo {title} {{Ferroelectricity from spin supercurrents in
  LiCuVO${}_{4}$}},}\ }\href {\doibase 10.1103/PhysRevB.83.100409} {\bibfield
  {journal} {\bibinfo  {journal} {Phys. Rev. B}\ }\textbf {\bibinfo {volume}
  {83}},\ \bibinfo {pages} {100409} (\bibinfo {year} {2011})}\BibitemShut
  {NoStop}%
\bibitem [{\citenamefont {Seki}\ \emph {et~al.}(2008)\citenamefont {Seki},
  \citenamefont {Yamasaki}, \citenamefont {Soda}, \citenamefont {Matsuura},
  \citenamefont {Hirota},\ and\ \citenamefont
  {Tokura}}]{Seki_PRL_2008_LiCu2O2chiralitySwitching}%
  \BibitemOpen
  \bibfield  {author} {\bibinfo {author} {\bibfnamefont {S.}~\bibnamefont
  {Seki}}, \bibinfo {author} {\bibfnamefont {Y.}~\bibnamefont {Yamasaki}},
  \bibinfo {author} {\bibfnamefont {M.}~\bibnamefont {Soda}}, \bibinfo {author}
  {\bibfnamefont {M.}~\bibnamefont {Matsuura}}, \bibinfo {author}
  {\bibfnamefont {K.}~\bibnamefont {Hirota}}, \ and\ \bibinfo {author}
  {\bibfnamefont {Y.}~\bibnamefont {Tokura}},\ }\bibfield  {title} {\enquote
  {\bibinfo {title} {{Correlation between Spin Helicity and an Electric
  Polarization Vector in Quantum-Spin Chain Magnet
  ${\mathrm{LiCu}}_{2}{\mathrm{O}}_{2}$}},}\ }\href {\doibase
  10.1103/PhysRevLett.100.127201} {\bibfield  {journal} {\bibinfo  {journal}
  {Phys. Rev. Lett.}\ }\textbf {\bibinfo {volume} {100}},\ \bibinfo {pages}
  {127201} (\bibinfo {year} {2008})}\BibitemShut {NoStop}%
\bibitem [{\citenamefont {Furukawa}\ \emph {et~al.}(2010)\citenamefont
  {Furukawa}, \citenamefont {Sato},\ and\ \citenamefont
  {Onoda}}]{Furukawa_PRL_2010_1Delectromagnon}%
  \BibitemOpen
  \bibfield  {author} {\bibinfo {author} {\bibfnamefont {S.}~\bibnamefont
  {Furukawa}}, \bibinfo {author} {\bibfnamefont {M.}~\bibnamefont {Sato}}, \
  and\ \bibinfo {author} {\bibfnamefont {S.}~\bibnamefont {Onoda}},\ }\bibfield
   {title} {\enquote {\bibinfo {title} {{Chiral Order and Electromagnetic
  Dynamics in One-Dimensional Multiferroic Cuprates}},}\ }\href {\doibase
  10.1103/PhysRevLett.105.257205} {\bibfield  {journal} {\bibinfo  {journal}
  {Phys. Rev. Lett.}\ }\textbf {\bibinfo {volume} {105}},\ \bibinfo {pages}
  {257205} (\bibinfo {year} {2010})}\BibitemShut {NoStop}%
\bibitem [{\citenamefont {Katsura}\ \emph {et~al.}(2005)\citenamefont
  {Katsura}, \citenamefont {Nagaosa},\ and\ \citenamefont
  {Balatsky}}]{Katsura_PRL_2005_MFmicro}%
  \BibitemOpen
  \bibfield  {author} {\bibinfo {author} {\bibfnamefont {H.}~\bibnamefont
  {Katsura}}, \bibinfo {author} {\bibfnamefont {N.}~\bibnamefont {Nagaosa}}, \
  and\ \bibinfo {author} {\bibfnamefont {A.~V.}\ \bibnamefont {Balatsky}},\
  }\bibfield  {title} {\enquote {\bibinfo {title} {Spin current and
  magnetoelectric effect in noncollinear magnets},}\ }\href {\doibase
  10.1103/PhysRevLett.95.057205} {\bibfield  {journal} {\bibinfo  {journal}
  {Phys. Rev. Lett.}\ }\textbf {\bibinfo {volume} {95}},\ \bibinfo {pages}
  {057205} (\bibinfo {year} {2005})}\BibitemShut {NoStop}%
\bibitem [{\citenamefont {Mostovoy}(2006)}]{Mostovoy_PRL_2006_MFmacro}%
  \BibitemOpen
  \bibfield  {author} {\bibinfo {author} {\bibfnamefont {M.}~\bibnamefont
  {Mostovoy}},\ }\bibfield  {title} {\enquote {\bibinfo {title}
  {Ferroelectricity in spiral magnets},}\ }\href {\doibase
  10.1103/PhysRevLett.96.067601} {\bibfield  {journal} {\bibinfo  {journal}
  {Phys. Rev. Lett.}\ }\textbf {\bibinfo {volume} {96}},\ \bibinfo {pages}
  {067601} (\bibinfo {year} {2006})}\BibitemShut {NoStop}%
\bibitem [{\citenamefont {Baran}\ \emph {et~al.}(2006)\citenamefont {Baran},
  \citenamefont {Jedrzejczak}, \citenamefont {Szymczak}, \citenamefont
  {Maltsev}, \citenamefont {Kamieniarz}, \citenamefont {Szukowski},
  \citenamefont {Loison}, \citenamefont {Ormeci}, \citenamefont {Drechsler},\
  and\ \citenamefont {Rosner}}]{Baran_PhysStatSol_2006_LinariteInitiation}%
  \BibitemOpen
  \bibfield  {author} {\bibinfo {author} {\bibfnamefont {M.}~\bibnamefont
  {Baran}}, \bibinfo {author} {\bibfnamefont {A.}~\bibnamefont {Jedrzejczak}},
  \bibinfo {author} {\bibfnamefont {H.}~\bibnamefont {Szymczak}}, \bibinfo
  {author} {\bibfnamefont {V.}~\bibnamefont {Maltsev}}, \bibinfo {author}
  {\bibfnamefont {G.}~\bibnamefont {Kamieniarz}}, \bibinfo {author}
  {\bibfnamefont {G.}~\bibnamefont {Szukowski}}, \bibinfo {author}
  {\bibfnamefont {C.}~\bibnamefont {Loison}}, \bibinfo {author} {\bibfnamefont
  {A.}~\bibnamefont {Ormeci}}, \bibinfo {author} {\bibfnamefont {S.-L.}\
  \bibnamefont {Drechsler}}, \ and\ \bibinfo {author} {\bibfnamefont
  {H.}~\bibnamefont {Rosner}},\ }\bibfield  {title} {\enquote {\bibinfo {title}
  {{Quasi-one-dimensional $S = 1/2$ magnet Pb[Cu(SO$_4$(OH)$_2$]: frustration
  due to competing in-chain exchange}},}\ }\href {\doibase
  10.1002/pssc.200562523} {\bibfield  {journal} {\bibinfo  {journal} {Phys.
  Status Solidi C}\ }\textbf {\bibinfo {volume} {3}},\ \bibinfo {pages} {220}
  (\bibinfo {year} {2006})}\BibitemShut {NoStop}%
\bibitem [{\citenamefont {Sch\"apers}\ \emph {et~al.}(2013)\citenamefont
  {Sch\"apers}, \citenamefont {Wolter}, \citenamefont {Drechsler},
  \citenamefont {Nishimoto}, \citenamefont {M\"uller}, \citenamefont
  {Abdel-Hafiez}, \citenamefont {Schottenhamel}, \citenamefont {B\"uchner},
  \citenamefont {Richter}, \citenamefont {Ouladdiaf}, \citenamefont {Uhlarz},
  \citenamefont {Beyer}, \citenamefont {Skourski}, \citenamefont {Wosnitza},
  \citenamefont {Rule}, \citenamefont {Ryll}, \citenamefont {Klemke},
  \citenamefont {Kiefer}, \citenamefont {Reehuis}, \citenamefont {Willenberg},\
  and\ \citenamefont {S\"ullow}}]{Schapers_PRB_2013_LinariteBulk}%
  \BibitemOpen
  \bibfield  {author} {\bibinfo {author} {\bibfnamefont {M.}~\bibnamefont
  {Sch\"apers}}, \bibinfo {author} {\bibfnamefont {A.~U.~B.}\ \bibnamefont
  {Wolter}}, \bibinfo {author} {\bibfnamefont {S.-L.}\ \bibnamefont
  {Drechsler}}, \bibinfo {author} {\bibfnamefont {S.}~\bibnamefont
  {Nishimoto}}, \bibinfo {author} {\bibfnamefont {K.-H.}\ \bibnamefont
  {M\"uller}}, \bibinfo {author} {\bibfnamefont {M.}~\bibnamefont
  {Abdel-Hafiez}}, \bibinfo {author} {\bibfnamefont {W.}~\bibnamefont
  {Schottenhamel}}, \bibinfo {author} {\bibfnamefont {B.}~\bibnamefont
  {B\"uchner}}, \bibinfo {author} {\bibfnamefont {J.}~\bibnamefont {Richter}},
  \bibinfo {author} {\bibfnamefont {B.}~\bibnamefont {Ouladdiaf}}, \bibinfo
  {author} {\bibfnamefont {M.}~\bibnamefont {Uhlarz}}, \bibinfo {author}
  {\bibfnamefont {R.}~\bibnamefont {Beyer}}, \bibinfo {author} {\bibfnamefont
  {Y.}~\bibnamefont {Skourski}}, \bibinfo {author} {\bibfnamefont
  {J.}~\bibnamefont {Wosnitza}}, \bibinfo {author} {\bibfnamefont {K.~C.}\
  \bibnamefont {Rule}}, \bibinfo {author} {\bibfnamefont {H.}~\bibnamefont
  {Ryll}}, \bibinfo {author} {\bibfnamefont {B.}~\bibnamefont {Klemke}},
  \bibinfo {author} {\bibfnamefont {K.}~\bibnamefont {Kiefer}}, \bibinfo
  {author} {\bibfnamefont {M.}~\bibnamefont {Reehuis}}, \bibinfo {author}
  {\bibfnamefont {B.}~\bibnamefont {Willenberg}}, \ and\ \bibinfo {author}
  {\bibfnamefont {S.}~\bibnamefont {S\"ullow}},\ }\bibfield  {title} {\enquote
  {\bibinfo {title} {{Thermodynamic properties of the anisotropic frustrated
  spin-chain compound linarite PbCuSO${}_{4}$(OH)${}_{2}$}},}\ }\href {\doibase
  10.1103/PhysRevB.88.184410} {\bibfield  {journal} {\bibinfo  {journal} {Phys.
  Rev. B}\ }\textbf {\bibinfo {volume} {88}},\ \bibinfo {pages} {184410}
  (\bibinfo {year} {2013})}\BibitemShut {NoStop}%
\bibitem [{\citenamefont {Willenberg}\ \emph {et~al.}(2016)\citenamefont
  {Willenberg}, \citenamefont {Sch\"apers}, \citenamefont {Wolter},
  \citenamefont {Drechsler}, \citenamefont {Reehuis}, \citenamefont {Hoffmann},
  \citenamefont {B\"uchner}, \citenamefont {Studer}, \citenamefont {Rule},
  \citenamefont {Ouladdiaf}, \citenamefont {S\"ullow},\ and\ \citenamefont
  {Nishimoto}}]{Willenberg_PRL_2016_LinariteSDWs}%
  \BibitemOpen
  \bibfield  {author} {\bibinfo {author} {\bibfnamefont {B.}~\bibnamefont
  {Willenberg}}, \bibinfo {author} {\bibfnamefont {M.}~\bibnamefont
  {Sch\"apers}}, \bibinfo {author} {\bibfnamefont {A.~U.~B.}\ \bibnamefont
  {Wolter}}, \bibinfo {author} {\bibfnamefont {S.-L.}\ \bibnamefont
  {Drechsler}}, \bibinfo {author} {\bibfnamefont {M.}~\bibnamefont {Reehuis}},
  \bibinfo {author} {\bibfnamefont {J.-U.}\ \bibnamefont {Hoffmann}}, \bibinfo
  {author} {\bibfnamefont {B.}~\bibnamefont {B\"uchner}}, \bibinfo {author}
  {\bibfnamefont {A.~J.}\ \bibnamefont {Studer}}, \bibinfo {author}
  {\bibfnamefont {K.~C.}\ \bibnamefont {Rule}}, \bibinfo {author}
  {\bibfnamefont {B.}~\bibnamefont {Ouladdiaf}}, \bibinfo {author}
  {\bibfnamefont {S.}~\bibnamefont {S\"ullow}}, \ and\ \bibinfo {author}
  {\bibfnamefont {S.}~\bibnamefont {Nishimoto}},\ }\bibfield  {title} {\enquote
  {\bibinfo {title} {{Complex Field-Induced States in Linarite
  ${\mathrm{PbCuSO}}_{4}(\mathrm{OH}{)}_{2}$ with a Variety of High-Order
  Exotic Spin-Density Wave States}},}\ }\href {\doibase
  10.1103/PhysRevLett.116.047202} {\bibfield  {journal} {\bibinfo  {journal}
  {Phys. Rev. Lett.}\ }\textbf {\bibinfo {volume} {116}},\ \bibinfo {pages}
  {047202} (\bibinfo {year} {2016})}\BibitemShut {NoStop}%
\bibitem [{\citenamefont {Yasui}\ \emph {et~al.}(2011)\citenamefont {Yasui},
  \citenamefont {Sato},\ and\ \citenamefont
  {Terasaki}}]{Yasui_JPSJ_2011_LinariteMF}%
  \BibitemOpen
  \bibfield  {author} {\bibinfo {author} {\bibfnamefont {Y.}~\bibnamefont
  {Yasui}}, \bibinfo {author} {\bibfnamefont {M.}~\bibnamefont {Sato}}, \ and\
  \bibinfo {author} {\bibfnamefont {I.}~\bibnamefont {Terasaki}},\ }\bibfield
  {title} {\enquote {\bibinfo {title} {{Multiferroic Behavior in the
  Quasi-One-Dimensional Frustrated Spin-1/2 System PbCuSO$_4$(OH)$_2$ with
  CuO$_2$ Ribbon Chains}},}\ }\href {\doibase 10.1143/JPSJ.80.033707}
  {\bibfield  {journal} {\bibinfo  {journal} {J. Phys. Soc. Jap.}\ }\textbf
  {\bibinfo {volume} {80}},\ \bibinfo {pages} {033707} (\bibinfo {year}
  {2011})}\BibitemShut {NoStop}%
\bibitem [{\citenamefont {Willenberg}\ \emph {et~al.}(2012)\citenamefont
  {Willenberg}, \citenamefont {Sch\"apers}, \citenamefont {Rule}, \citenamefont
  {S\"ullow}, \citenamefont {Reehuis}, \citenamefont {Ryll}, \citenamefont
  {Klemke}, \citenamefont {Kiefer}, \citenamefont {Schottenhamel},
  \citenamefont {B\"uchner}, \citenamefont {Ouladdiaf}, \citenamefont {Uhlarz},
  \citenamefont {Beyer}, \citenamefont {Wosnitza},\ and\ \citenamefont
  {Wolter}}]{Willenberg_PRL_2012_LinariteFrustrated}%
  \BibitemOpen
  \bibfield  {author} {\bibinfo {author} {\bibfnamefont {B.}~\bibnamefont
  {Willenberg}}, \bibinfo {author} {\bibfnamefont {M.}~\bibnamefont
  {Sch\"apers}}, \bibinfo {author} {\bibfnamefont {K.~C.}\ \bibnamefont
  {Rule}}, \bibinfo {author} {\bibfnamefont {S.}~\bibnamefont {S\"ullow}},
  \bibinfo {author} {\bibfnamefont {M.}~\bibnamefont {Reehuis}}, \bibinfo
  {author} {\bibfnamefont {H.}~\bibnamefont {Ryll}}, \bibinfo {author}
  {\bibfnamefont {B.}~\bibnamefont {Klemke}}, \bibinfo {author} {\bibfnamefont
  {K.}~\bibnamefont {Kiefer}}, \bibinfo {author} {\bibfnamefont
  {W.}~\bibnamefont {Schottenhamel}}, \bibinfo {author} {\bibfnamefont
  {B.}~\bibnamefont {B\"uchner}}, \bibinfo {author} {\bibfnamefont
  {B.}~\bibnamefont {Ouladdiaf}}, \bibinfo {author} {\bibfnamefont
  {M.}~\bibnamefont {Uhlarz}}, \bibinfo {author} {\bibfnamefont
  {R.}~\bibnamefont {Beyer}}, \bibinfo {author} {\bibfnamefont
  {J.}~\bibnamefont {Wosnitza}}, \ and\ \bibinfo {author} {\bibfnamefont
  {A.~U.~B.}\ \bibnamefont {Wolter}},\ }\bibfield  {title} {\enquote {\bibinfo
  {title} {{Magnetic Frustration in a Quantum Spin Chain: The Case of Linarite
  ${\mathrm{PbCuSO}}_{4}(\mathrm{OH}{)}_{2}$}},}\ }\href {\doibase
  10.1103/PhysRevLett.108.117202} {\bibfield  {journal} {\bibinfo  {journal}
  {Phys. Rev. Lett.}\ }\textbf {\bibinfo {volume} {108}},\ \bibinfo {pages}
  {117202} (\bibinfo {year} {2012})}\BibitemShut {NoStop}%
\bibitem [{\citenamefont {Wolter}\ \emph {et~al.}(2012)\citenamefont {Wolter},
  \citenamefont {Lipps}, \citenamefont {Sch\"apers}, \citenamefont {Drechsler},
  \citenamefont {Nishimoto}, \citenamefont {Vogel}, \citenamefont {Kataev},
  \citenamefont {B\"uchner}, \citenamefont {Rosner}, \citenamefont {Schmitt},
  \citenamefont {Uhlarz}, \citenamefont {Skourski}, \citenamefont {Wosnitza},
  \citenamefont {S\"ullow},\ and\ \citenamefont
  {Rule}}]{WolterLipps_PRB_2012_LinariteESR}%
  \BibitemOpen
  \bibfield  {author} {\bibinfo {author} {\bibfnamefont {A.~U.~B.}\
  \bibnamefont {Wolter}}, \bibinfo {author} {\bibfnamefont {F.}~\bibnamefont
  {Lipps}}, \bibinfo {author} {\bibfnamefont {M.}~\bibnamefont {Sch\"apers}},
  \bibinfo {author} {\bibfnamefont {S.-L.}\ \bibnamefont {Drechsler}}, \bibinfo
  {author} {\bibfnamefont {S.}~\bibnamefont {Nishimoto}}, \bibinfo {author}
  {\bibfnamefont {R.}~\bibnamefont {Vogel}}, \bibinfo {author} {\bibfnamefont
  {V.}~\bibnamefont {Kataev}}, \bibinfo {author} {\bibfnamefont
  {B.}~\bibnamefont {B\"uchner}}, \bibinfo {author} {\bibfnamefont
  {H.}~\bibnamefont {Rosner}}, \bibinfo {author} {\bibfnamefont
  {M.}~\bibnamefont {Schmitt}}, \bibinfo {author} {\bibfnamefont
  {M.}~\bibnamefont {Uhlarz}}, \bibinfo {author} {\bibfnamefont
  {Y.}~\bibnamefont {Skourski}}, \bibinfo {author} {\bibfnamefont
  {J.}~\bibnamefont {Wosnitza}}, \bibinfo {author} {\bibfnamefont
  {S.}~\bibnamefont {S\"ullow}}, \ and\ \bibinfo {author} {\bibfnamefont
  {K.~C.}\ \bibnamefont {Rule}},\ }\bibfield  {title} {\enquote {\bibinfo
  {title} {{Magnetic properties and exchange integrals of the frustrated chain
  cuprate linarite PbCuSO${}_{4}$(OH)${}_{2}$}},}\ }\href {\doibase
  10.1103/PhysRevB.85.014407} {\bibfield  {journal} {\bibinfo  {journal} {Phys.
  Rev. B}\ }\textbf {\bibinfo {volume} {85}},\ \bibinfo {pages} {014407}
  (\bibinfo {year} {2012})}\BibitemShut {NoStop}%
\bibitem [{\citenamefont {Bachmann}\ and\ \citenamefont
  {Zemann}(1961)}]{Bachmann_ActCryst_1961_LinariteStructure}%
  \BibitemOpen
  \bibfield  {author} {\bibinfo {author} {\bibfnamefont {H.~G.}\ \bibnamefont
  {Bachmann}}\ and\ \bibinfo {author} {\bibfnamefont {J.}~\bibnamefont
  {Zemann}},\ }\bibfield  {title} {\enquote {\bibinfo {title} {{Die
  Kristallstruktur von Linarit PbCuSO4(OH)2}},}\ }\href {\doibase
  10.1107/S0365110X61002254} {\bibfield  {journal} {\bibinfo  {journal} {Acta
  Crystallographica}\ }\textbf {\bibinfo {volume} {14}},\ \bibinfo {pages}
  {747} (\bibinfo {year} {1961})}\BibitemShut {NoStop}%
\bibitem [{Note1()}]{Note1}%
  \BibitemOpen
  \bibinfo {note} {Linarite crystals typically have a prismatic morphology with
  elongation along the $b$ axis. In addition to this easily identifiable
  direction, typical prism also has the largest facet corresponding to [100]
  plane (the main cleavage plane of linarite). However, there is a reasonable
  chance of finding a crystal which has the [001] plane (the second cleavage
  plane) as the largest facet instead. These two cases were distinguished at
  the next step, where the x-ray diffraction check was performed.}\BibitemShut
  {Stop}%
\bibitem [{Note2()}]{Note2}%
  \BibitemOpen
  \bibinfo {note} {As in the standard reciprocal lattice notation.}\BibitemShut
  {Stop}%
\bibitem [{\citenamefont {Sch\"apers}\ \emph {et~al.}(2014)\citenamefont
  {Sch\"apers}, \citenamefont {Rosner}, \citenamefont {Drechsler},
  \citenamefont {S\"ullow}, \citenamefont {Vogel}, \citenamefont {B\"uchner},\
  and\ \citenamefont {Wolter}}]{Schapers_PRB_2014_LinariteNMR}%
  \BibitemOpen
  \bibfield  {author} {\bibinfo {author} {\bibfnamefont {M.}~\bibnamefont
  {Sch\"apers}}, \bibinfo {author} {\bibfnamefont {H.}~\bibnamefont {Rosner}},
  \bibinfo {author} {\bibfnamefont {S.-L.}\ \bibnamefont {Drechsler}}, \bibinfo
  {author} {\bibfnamefont {S.}~\bibnamefont {S\"ullow}}, \bibinfo {author}
  {\bibfnamefont {R.}~\bibnamefont {Vogel}}, \bibinfo {author} {\bibfnamefont
  {B.}~\bibnamefont {B\"uchner}}, \ and\ \bibinfo {author} {\bibfnamefont
  {A.~U.~B.}\ \bibnamefont {Wolter}},\ }\bibfield  {title} {\enquote {\bibinfo
  {title} {{Magnetic and electronic structure of the frustrated spin-chain
  compound linarite ${\mathrm{PbCuSO}}_{4}(\mathrm{OH}){}_{2}$}},}\ }\href
  {\doibase 10.1103/PhysRevB.90.224417} {\bibfield  {journal} {\bibinfo
  {journal} {Phys. Rev. B}\ }\textbf {\bibinfo {volume} {90}},\ \bibinfo
  {pages} {224417} (\bibinfo {year} {2014})}\BibitemShut {NoStop}%
\bibitem [{\citenamefont {Dzyaloshinskii}(1960)}]{Dzyaloshinskii_JETP_1959_ME}%
  \BibitemOpen
  \bibfield  {author} {\bibinfo {author} {\bibfnamefont {I.~E.}\ \bibnamefont
  {Dzyaloshinskii}},\ }\bibfield  {title} {\enquote {\bibinfo {title}
  {{Magnetoelectric Effect in Chromium Oxide}},}\ }\href
  {http://www.jetp.ac.ru/cgi-bin/e/index/e/10/3/p628?a=list} {\bibfield
  {journal} {\bibinfo  {journal} {Sov. Phys. JETP}\ }\textbf {\bibinfo {volume}
  {10}},\ \bibinfo {pages} {628} (\bibinfo {year} {1960})}\BibitemShut
  {NoStop}%
\bibitem [{\citenamefont {Astrov}(1960)}]{Astrov_JETP_1960_ME1}%
  \BibitemOpen
  \bibfield  {author} {\bibinfo {author} {\bibfnamefont {D.~N.}\ \bibnamefont
  {Astrov}},\ }\bibfield  {title} {\enquote {\bibinfo {title} {{The
  Magnetoelectric Effect in Antiferromagnetics}},}\ }\href
  {http://www.jetp.ac.ru/cgi-bin/e/index/e/11/3/p708?a=list} {\bibfield
  {journal} {\bibinfo  {journal} {Sov. Phys. JETP}\ }\textbf {\bibinfo {volume}
  {11}},\ \bibinfo {pages} {708} (\bibinfo {year} {1960})}\BibitemShut
  {NoStop}%
\bibitem [{\citenamefont {Astrov}(1961)}]{Astrov_JETP_1961_ME2}%
  \BibitemOpen
  \bibfield  {author} {\bibinfo {author} {\bibfnamefont {D.~N.}\ \bibnamefont
  {Astrov}},\ }\bibfield  {title} {\enquote {\bibinfo {title} {{Magnetoelectric
  Effect in Chromium Oxide}},}\ }\href
  {http://www.jetp.ac.ru/cgi-bin/e/index/e/13/4/p729?a=list} {\bibfield
  {journal} {\bibinfo  {journal} {Sov. Phys. JETP}\ }\textbf {\bibinfo {volume}
  {13}},\ \bibinfo {pages} {729} (\bibinfo {year} {1961})}\BibitemShut
  {NoStop}%
\bibitem [{\citenamefont
  {Dvo{\v{r}}{\'a}k}(1974)}]{Dvorak_Ferr_1974_improperReview}%
  \BibitemOpen
  \bibfield  {author} {\bibinfo {author} {\bibfnamefont {V.}~\bibnamefont
  {Dvo{\v{r}}{\'a}k}},\ }\bibfield  {title} {\enquote {\bibinfo {title}
  {Improper ferroelectrics},}\ }\href
  {http://www.tandfonline.com/doi/abs/10.1080/00150197408237942} {\bibfield
  {journal} {\bibinfo  {journal} {Ferroelectrics}\ }\textbf {\bibinfo {volume}
  {7}},\ \bibinfo {pages} {1} (\bibinfo {year} {1974})}\BibitemShut {NoStop}%
\bibitem [{\citenamefont {H\"alg}\ \emph {et~al.}(2015)\citenamefont {H\"alg},
  \citenamefont {H\"uvonen}, \citenamefont {Guidi}, \citenamefont
  {Quintero-Castro}, \citenamefont {Boehm}, \citenamefont {Regnault},
  \citenamefont {Hagiwara},\ and\ \citenamefont
  {Zheludev}}]{HaelgHuvonen_PRB_2015_NTENPscaling}%
  \BibitemOpen
  \bibfield  {author} {\bibinfo {author} {\bibfnamefont {M.}~\bibnamefont
  {H\"alg}}, \bibinfo {author} {\bibfnamefont {D.}~\bibnamefont {H\"uvonen}},
  \bibinfo {author} {\bibfnamefont {T.}~\bibnamefont {Guidi}}, \bibinfo
  {author} {\bibfnamefont {D.~L.}\ \bibnamefont {Quintero-Castro}}, \bibinfo
  {author} {\bibfnamefont {M.}~\bibnamefont {Boehm}}, \bibinfo {author}
  {\bibfnamefont {L.~P.}\ \bibnamefont {Regnault}}, \bibinfo {author}
  {\bibfnamefont {M.}~\bibnamefont {Hagiwara}}, \ and\ \bibinfo {author}
  {\bibfnamefont {A.}~\bibnamefont {Zheludev}},\ }\bibfield  {title} {\enquote
  {\bibinfo {title} {{Finite-temperature scaling of spin correlations in an
  experimental realization of the one-dimensional Ising quantum critical
  point}},}\ }\href {\doibase 10.1103/PhysRevB.92.014412} {\bibfield  {journal}
  {\bibinfo  {journal} {Phys. Rev. B}\ }\textbf {\bibinfo {volume} {92}},\
  \bibinfo {pages} {014412} (\bibinfo {year} {2015})}\BibitemShut {NoStop}%
\bibitem [{\citenamefont {Kenzelmann}\ \emph {et~al.}(2007)\citenamefont
  {Kenzelmann}, \citenamefont {Lawes}, \citenamefont {Harris}, \citenamefont
  {Gasparovic}, \citenamefont {Broholm}, \citenamefont {Ramirez}, \citenamefont
  {Jorge}, \citenamefont {Jaime}, \citenamefont {Park}, \citenamefont {Huang},
  \citenamefont {Shapiro},\ and\ \citenamefont
  {Demianets}}]{Kenzelmann_PRL_2007_TriangularFerro}%
  \BibitemOpen
  \bibfield  {author} {\bibinfo {author} {\bibfnamefont {M.}~\bibnamefont
  {Kenzelmann}}, \bibinfo {author} {\bibfnamefont {G.}~\bibnamefont {Lawes}},
  \bibinfo {author} {\bibfnamefont {A.~B.}\ \bibnamefont {Harris}}, \bibinfo
  {author} {\bibfnamefont {G.}~\bibnamefont {Gasparovic}}, \bibinfo {author}
  {\bibfnamefont {C.}~\bibnamefont {Broholm}}, \bibinfo {author} {\bibfnamefont
  {A.~P.}\ \bibnamefont {Ramirez}}, \bibinfo {author} {\bibfnamefont {G.~A.}\
  \bibnamefont {Jorge}}, \bibinfo {author} {\bibfnamefont {M.}~\bibnamefont
  {Jaime}}, \bibinfo {author} {\bibfnamefont {S.}~\bibnamefont {Park}},
  \bibinfo {author} {\bibfnamefont {Q.}~\bibnamefont {Huang}}, \bibinfo
  {author} {\bibfnamefont {A.~Ya.}\ \bibnamefont {Shapiro}}, \ and\ \bibinfo
  {author} {\bibfnamefont {L.~A.}\ \bibnamefont {Demianets}},\ }\bibfield
  {title} {\enquote {\bibinfo {title} {Direct transition from a disordered to a
  multiferroic phase on a triangular lattice},}\ }\href {\doibase
  10.1103/PhysRevLett.98.267205} {\bibfield  {journal} {\bibinfo  {journal}
  {Phys. Rev. Lett.}\ }\textbf {\bibinfo {volume} {98}},\ \bibinfo {pages}
  {267205} (\bibinfo {year} {2007})}\BibitemShut {NoStop}%
\bibitem [{\citenamefont {Harris}(2007)}]{Harris_PRB_2007_MFlandau}%
  \BibitemOpen
  \bibfield  {author} {\bibinfo {author} {\bibfnamefont {A.~B.}\ \bibnamefont
  {Harris}},\ }\bibfield  {title} {\enquote {\bibinfo {title} {{Landau analysis
  of the symmetry of the magnetic structure and magnetoelectric interaction in
  multiferroics}},}\ }\href {\doibase 10.1103/PhysRevB.76.054447} {\bibfield
  {journal} {\bibinfo  {journal} {Phys. Rev. B}\ }\textbf {\bibinfo {volume}
  {76}},\ \bibinfo {pages} {054447} (\bibinfo {year} {2007})}\BibitemShut
  {NoStop}%
\bibitem [{\citenamefont {M\"uhlbauer}\ \emph {et~al.}(2011)\citenamefont
  {M\"uhlbauer}, \citenamefont {Gvasaliya}, \citenamefont {Pomjakushina},\ and\
  \citenamefont {Zheludev}}]{Muehlbauer_PRB_2011_BaCuGeOmultiK}%
  \BibitemOpen
  \bibfield  {author} {\bibinfo {author} {\bibfnamefont {S.}~\bibnamefont
  {M\"uhlbauer}}, \bibinfo {author} {\bibfnamefont {S.~N.}\ \bibnamefont
  {Gvasaliya}}, \bibinfo {author} {\bibfnamefont {E.}~\bibnamefont
  {Pomjakushina}}, \ and\ \bibinfo {author} {\bibfnamefont {A.}~\bibnamefont
  {Zheludev}},\ }\bibfield  {title} {\enquote {\bibinfo {title} {{Double-$k$
  phase of the Dzyaloshinskii-Moriya helimagnet
  Ba${}_{2}$CuGe${}_{2}$O${}_{7}$}},}\ }\href {\doibase
  10.1103/PhysRevB.84.180406} {\bibfield  {journal} {\bibinfo  {journal} {Phys.
  Rev. B}\ }\textbf {\bibinfo {volume} {84}},\ \bibinfo {pages} {180406}
  (\bibinfo {year} {2011})}\BibitemShut {NoStop}%
\bibitem [{\citenamefont
  {Dzyaloshinsky}(1958)}]{Dzyaloshinskii_JChemPhysSol_1958_DM}%
  \BibitemOpen
  \bibfield  {author} {\bibinfo {author} {\bibfnamefont {I.}~\bibnamefont
  {Dzyaloshinsky}},\ }\bibfield  {title} {\enquote {\bibinfo {title} {{A
  thermodynamic theory of 'weak' ferromagnetism of antiferromagnetics}},}\
  }\href {\doibase 10.1016/0022-3697(58)90076-3} {\bibfield  {journal}
  {\bibinfo  {journal} {J. Phys. Chem. Solids}\ }\textbf {\bibinfo {volume}
  {4}},\ \bibinfo {pages} {241} (\bibinfo {year} {1958})}\BibitemShut {NoStop}%
\bibitem [{\citenamefont {Moriya}(1960)}]{Moriya_PR_1960_DM}%
  \BibitemOpen
  \bibfield  {author} {\bibinfo {author} {\bibfnamefont {T.}~\bibnamefont
  {Moriya}},\ }\bibfield  {title} {\enquote {\bibinfo {title} {{Anisotropic
  Superexchange Interaction and Weak Ferromagnetism}},}\ }\href {\doibase
  10.1103/PhysRev.120.91} {\bibfield  {journal} {\bibinfo  {journal} {Phys.
  Rev.}\ }\textbf {\bibinfo {volume} {120}},\ \bibinfo {pages} {91} (\bibinfo
  {year} {1960})}\BibitemShut {NoStop}%
\bibitem [{Note3()}]{Note3}%
  \BibitemOpen
  \bibinfo {note} {Nonetheless, we note that the last non-featureless specific
  heat curve in the corresponding panel of Fig.~9 in Ref.~\cite
  {Schapers_PRB_2013_LinariteBulk} may give some room for
  speculation.}\BibitemShut {Stop}%
\end{thebibliography}%
\end{document}